\documentclass[english,journal=ancac3,layout=twocolumn,manuscript=article]{achemso}
\usepackage[T1]{fontenc}
\usepackage[utf8]{inputenc}   
\usepackage{color}
\usepackage{babel}
\usepackage{textcomp}
\usepackage{amssymb}
\usepackage{graphicx}
\usepackage{bm}
\usepackage{hyperref}
\usepackage{dcolumn}
\usepackage{mciteplus}
\usepackage{amsmath}
\usepackage{physics}
\usepackage{tikz}
\usepackage{standalone}

\makeatletter

\title{Selective Biexciton Generation Under Energy-Time Entangled Quantum Light in Quantum Dots}

\author{Kaiyue Peng}

\affiliation{Department of Chemistry, University of California, Berkeley, California 94720, USA}

\email{kaiyue_peng@berkeley.edu}

\author{Chieh Tsao}

\affiliation{Department of Chemistry, University of California, Berkeley, California 94720, USA}

\author{Hendrik Utzat}

\affiliation{Department of Chemistry, University of California, Berkeley, California 94720, USA}

\alsoaffiliation{Materials Sciences Division, Lawrence Berkeley National Laboratory, Berkeley, California 94720, USA}

\email{hutzat@berkeley.edu}

\author{Eran Rabani}

\affiliation{Fritz Haber Center for Molecular Dynamics, Institute of Chemistry and Institute of Applied Physics, Hebrew University of Jerusalem, Jerusalem 91904, Israel}

\alsoaffiliation{Department of Chemistry, University of California, Berkeley, California 94720, USA}

\alsoaffiliation{Materials Sciences Division, Lawrence Berkeley National Laboratory, Berkeley, California 94720, USA}

\email{eran.rabani@mail.huji.ac.il}


\setkeys{acs}{articletitle = true}
\usepackage{bm}
\AbstractOn
\SectionsOn

\makeatother
\begin{document}

\begin{abstract}
Quantum light provides opportunities for controlling multiphoton absorption beyond classical limits. Here, we investigate biexciton generation in nanocrystal quantum dots driven by energy-time-entangled photon pairs generated via spontaneous parametric down-conversion. We show that frequency-time correlations between pairs of photons increase the population of biexcitons over excitons, thereby lending specificity to the excitation of many-body states. By employing a three-level model, we demonstrate that biexciton generation depends nontrivially on the photon entanglement time and the pump bandwidth. We find that maximizing efficiency requires an optimally shaped entangled photon field rather than simply scaling parameters for a monotonic improvement. Extending to a realistic CdSe/CdS core-shell quantum dot containing many excitonic states coupled to the quantum field, we demonstrate that increasing the bi-photon arrival-time entanglement (closer arrival time) enhances constructive pathway interference and expands accessible excitation channels while preserving better energy-conservation excitation than classical light when generating biexcitons. Furthermore, tuning the time correlation properties enables selective excitation of closely spaced biexciton states. These results establish entangled photons as a tool for selective excitation and control of higher-order excited states in quantum-confined systems.
\end{abstract}

\section{Introduction}
Semiconductor quantum dots (QDs) have emerged as a versatile platform for optical applications, including light-emitting devices, lasers, and quantum information processing~\cite{arakawa2020progress,almutlaq2024engineering,bhattacharya2004quantum}. With their discrete energy spectra~\cite{empedocles1996photoluminescence,cho2023temperature,li2025exciton}, tunable band gaps~\cite{dressler2026phase,garcia2021semiconductor}, and strong light-matter interactions,~\cite{najer2019gated,empedocles1996photoluminescence,cho2023temperature,peng2023polaritonic,peng2024polariton} QDs offer an ideal platform for enabling precise control over optical absorption and emission processes~\cite{lodahl2004controlling,ropp2013nanoscale,huang2021bright}.
Crucially, they serve as robust, on-demand sources of both single photons and entangled photon pairs~\cite{schimpf2021quantum,zeeshan2019proposed,ding2024quantum}. In this context, the multi-exciton (MX) cascade--where a biexciton sequentially decays through an intermediate single-exciton state--plays a central role. Researchers have extensively leveraged this cascade in epitaxial QDs, and increasingly in colloidal QDs, to generate polarization- and time-bin-entangled photon pairs~\cite{nozik2008multiple,smith2013multiple,melnychuk2021multicarrier}. Consequently, the field has meticulously studied how physical parameters such as phonons, dephasing, and exciton fine-structure splitting impact the fidelity of the emitted quantum light~\cite{gywat2002biexcitons,emara2022enhancement,liu2025quantum}.

However, while the emission of entangled photons via the multi-exciton cascade is a well-established area of research, the inverse process -- driving the multi-exciton cascade upward using entangled quantum light -- remains essentially unexplored and presents a state-selective approach in the field of quantum spectroscopy.~\cite{tsao2025enhancing,eshun2021investigations,alvarez2025correlated} Traditionally, generating biexcitons relies on classical two-photon absorption (TPA). Because classical driving sources follow thermal or Poissonian statistics, driving such TPA process requires high peak intensities and yields stochastic multiphoton absorption events~\cite{kaldewey2017coherent}. Furthermore, this classical approach lacks intrinsic two-photon correlations. This absence restricts excitation selectivity and creates a non-negligible probability of simultaneously generating unwanted single excitons alongside the desired biexcitons.\cite{hu2020photoluminescence}

Emerging capabilities in quantum light spectroscopy, particularly the use of energy-time entangled photon pairs (typically generated via spontaneous parametric down-conversion, SPDC),\cite{gorlach2020quantum} present a significant opportunity to overcome these classical limitations. Such SPDC-generated photons exhibit strong temporal and spectral correlations and have been used in quantum communication and long-distance entanglement distribution~\cite{wengerowsky2018entanglement,tittel2000quantum}. Unlike classical light, entangled photons allow independent tuning of temporal correlations through the correlation bandwidth, without altering the overall spectral bandwidth. Importantly, while prior studies focused on two-photon absorption with entangled light via virtual states,~\cite{kang2020efficient, de2013role, dreano2023nonlinear} our work distinctly explores excitation through a real intermediate state. A strict time correlation ensures the photon pair arrives near-simultaneously, thereby ensuring the second absorption event occurs within the transient lifetime of the real intermediate single-exciton state. This additional degree of control is believed to allow enhancement of two-photon absorption at significantly lower photon flux, proposed to reduce background excitation and sample damage~\cite{burdick2021enhancing}, and inform the opposite process for the emission of entangled pairs.

Moreover, the interplay between bi-photon temporal correlations and energy conservation offers a mechanism to control multiphoton excitation pathways~\cite{gu2020manipulating,chen2022entangled}. Efficiently preparing a specific biexcitonic state is highly valuable for quantum-information applications and quantum-light generation,\cite{honig2014manifestation}  particularly when targeting the lowest-energy biexciton state involved in the biexciton--exciton radiative cascade.\cite{chen2001control,bounouar2015phonon} Previous work has focused on using such quantum light in stimulated linear absorption, non-linear photoelectron spectroscopy, pump-probe spectroscopies, and Raman techniques to demonstrate enhanced resolution~\cite{svidzinsky2021enhancing,dorfman2014stimulated,zhang2022entangled,schlawin2018entangled,gu2023photoelectron}. In contrast, the use of entangled photons to control multiexciton generation in realistic materials, such as QDs, remains less explored. In particular, a quantitative understanding of how photon correlations can be harnessed to achieve selective biexciton excitation and to predict performance in systems with complex excitonic structure is still missing.

In this work, we investigate how energy-time entangled quantum light generated via SPDC can be used to enhance and control biexciton generation in nanocrystal QDs through two-photon absorption via real intermediate fine-structure states. Using a minimal three-level model, we first demonstrate that the excitation of the biexciton via the intermediate exciton is significantly more efficient when the entangled light is optimally shaped. Specifically, by tuning the joint spectral amplitude, namely the pump bandwidth and the bi-photon entanglement time, we show that strong temporal entanglement enhances biexciton generation efficiency while suppressing singly excited state generation. We then generalize these intuitive findings using realistic atomistic simulations of CdSe/CdS core-shell QDs, confirming that this coherent control mechanism persists in complex systems of experimental relevance,~\cite{tsao2025heralded,alvarez2025correlated} while enabling highly selective excitation through the interplay of excitation pathway interference and energy conservation. These results suggest the possibility of selectively exciting many-body states via many-photon correlated states with appropriately prepared energy and time correlations. This "reverse" multi-exciton quantum cascade holds potential application in fundamental studies of many-body physics, including the limiting factors in entangled-pair generation via emission. The present results may also provide insights into the design of other quantum-light sources with tailored, nonfactorizable multiphoton spectral correlations for selective excitation and control of many-body states in complex materials.

\begin{figure}[t]
\centering
\includegraphics[width=8cm]{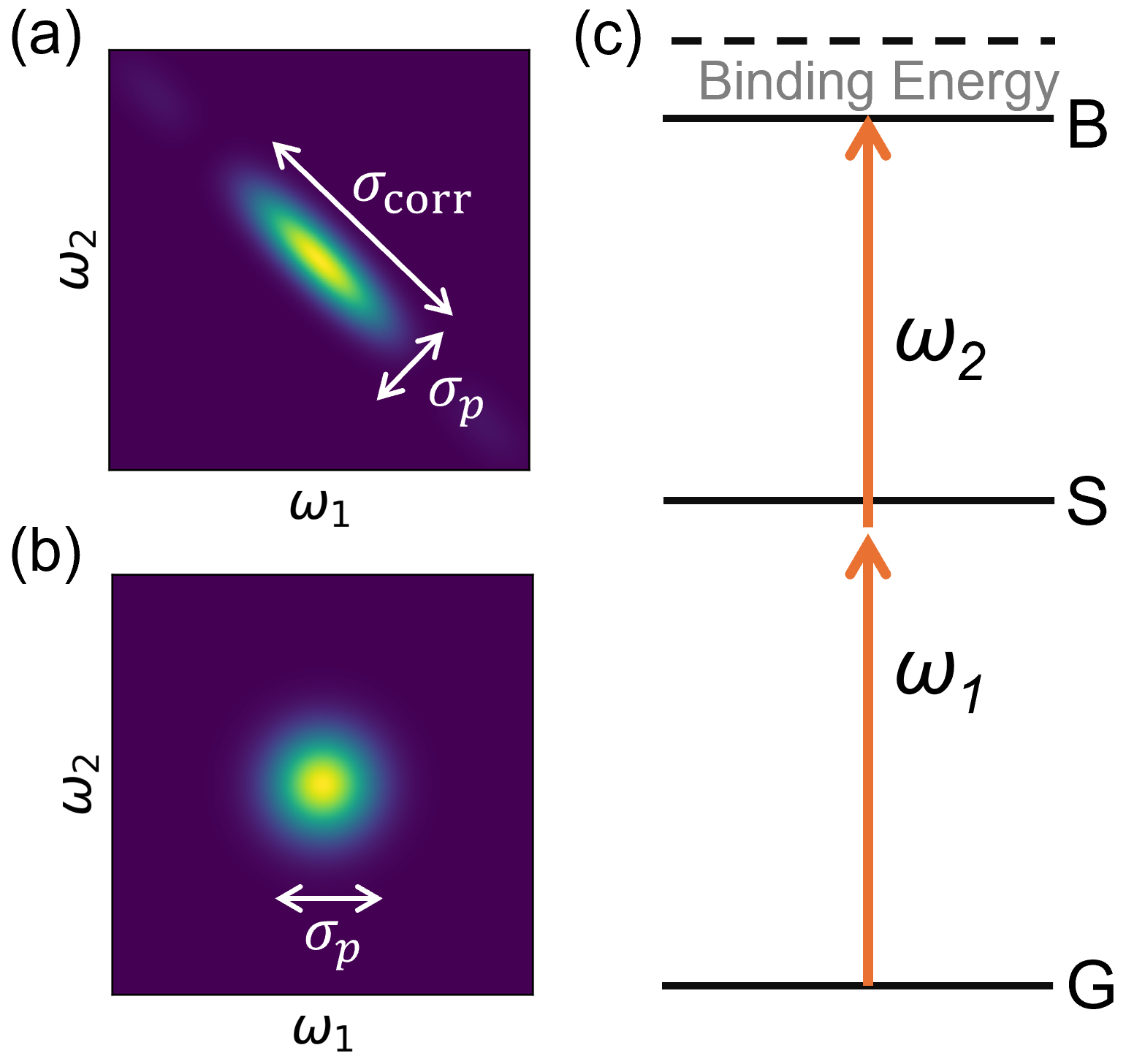}
\caption{Comparison of classical and quantum spectral correlations for $\omega_{1}^{0}=\omega_{2}^{0}$ for two photon absorption.  (a) The joint spectral intensity $\left|\Phi\left(\omega_{1},\omega_{2}\right)\right|$ for energy-time entangled photons. $\sigma_p$ is the pump bandwidth and $\sigma_\text{corr}$ is the correlation bandwidth. (b) The cross-product of classical Gaussian spectral profiles $|\psi(\omega_{1})\psi(\omega_{2})|$. (c) Schematic of the three-level ladder system $\left|g\right\rangle \rightarrow\left|S\right\rangle \rightarrow\left|B\right\rangle$ where excitonic state $\left|S\right\rangle $ serves as the intermediate pathway for biexciton generation.}
\label{fig:light_profile}
\end{figure}

\begin{figure*}[t]
\centering
\includegraphics[width=17cm]{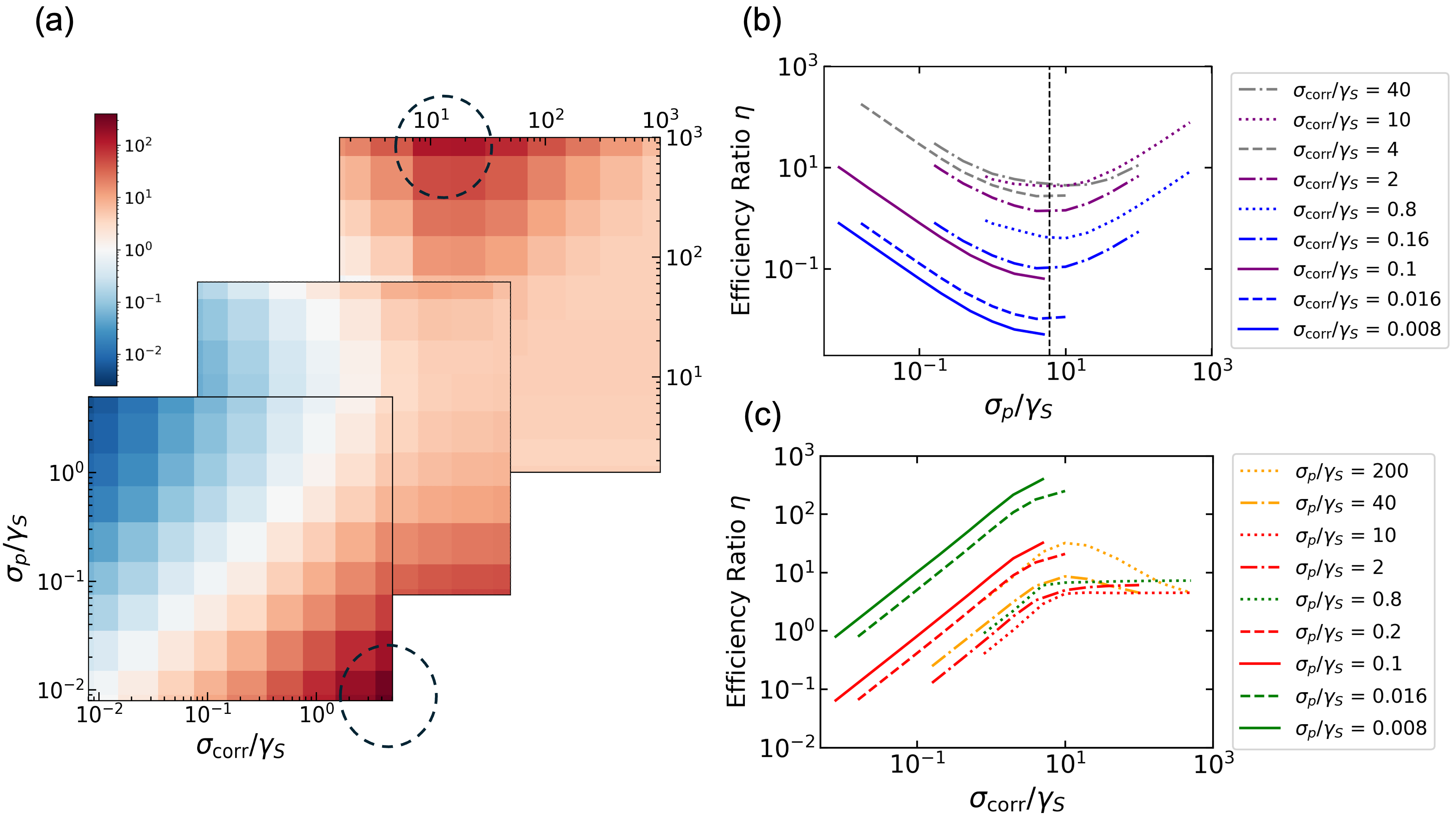}\caption{Energy--time entangled quantum light for selective biexciton generation in a three-level model. (a) Diagram of the efficiency ratio $\eta$ as a function of normalized correlation and pump bandwidth. The three square regions correspond to different excitonic dephasing rates, $\gamma_S$, and are aligned in normalized coordinates to illustrate the overall trend. The dashed circle highlights the regime of enhanced selective biexciton generation. (b) Dependence of efficiency ratio $\eta$ on normalized pump bandwidth $\frac{\sigma_{p}}{\gamma_{S}}$ for different normalized correlation bandwidths $\frac{\sigma_{\rm{corr}}}{\gamma_{S}}$. The vertical dashed line marks the approximate minimum at $\sigma_p/\gamma_S \approx 5$. (c) Dependence of the efficiency ratio $\eta$ on normalized correlation bandwidth $\frac{\sigma_{\rm{corr}}}{\gamma_{S}}$ for several normalized pump bandwidths $\frac{\sigma_{p}}{\gamma_{S}}$. In both panels (b) and (c), the several lines within each color family represent different dephasing rates $\hbar\gamma_S$ (ranging from 0.5 to 100 meV)}
\label{fig:3LS_analysis}
\end{figure*}

\begin{figure}[t]
\centering
\includegraphics[width=7cm]{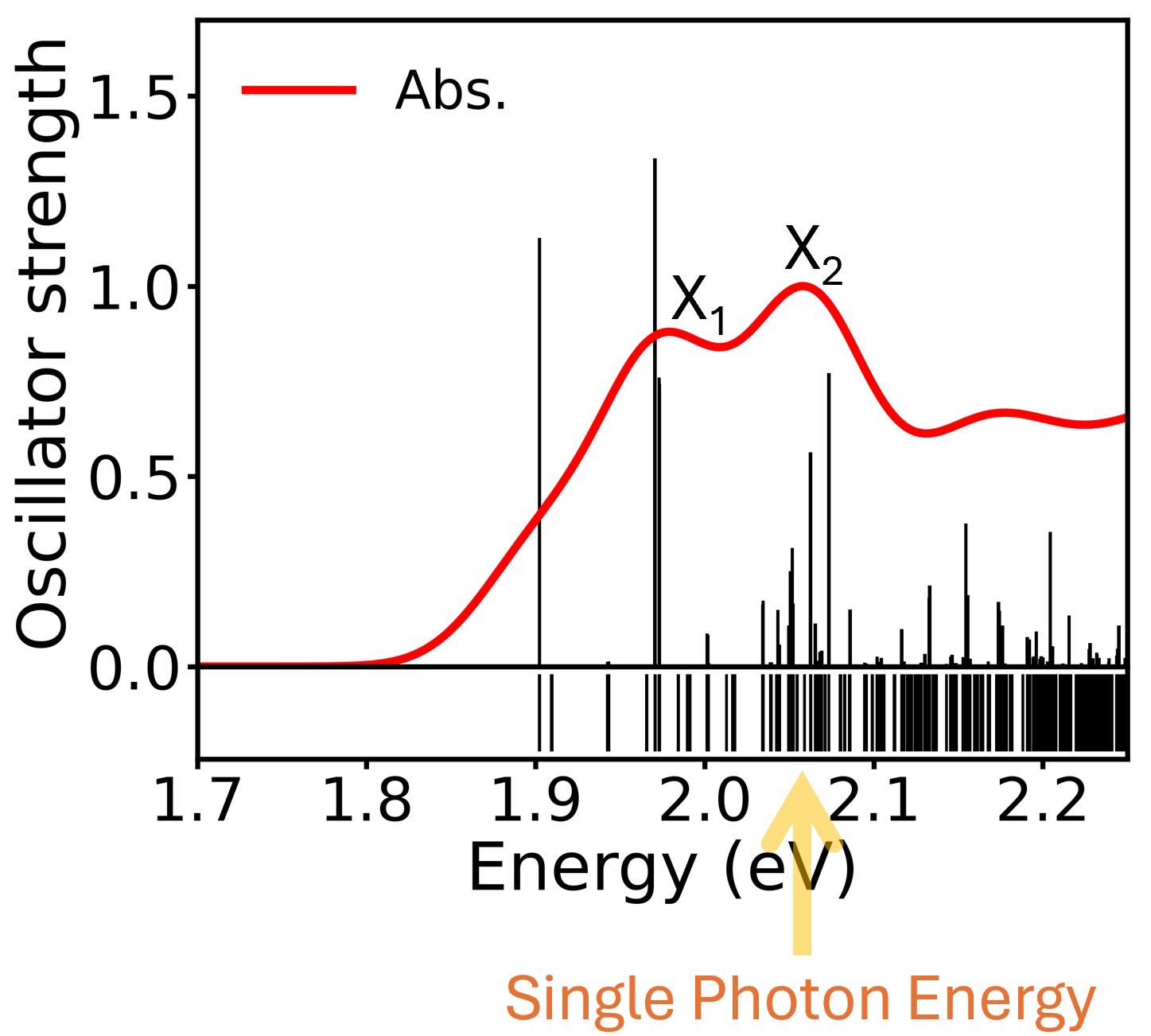}
\caption{Absorption spectrum of a $3.9$~nm diameter CdSe/$2$~ML CdS core-shell nanocrystal. The vertical lines in the upper subplot indicate the magnitude of the oscillator strength of the transition from the ground state to that excitonic state. The continuous absorption spectra (red curve) was obtained by broadening the individual transitions with a Gaussian function with a variance of $35$~meV. The yellow arrow indicates the bi-photon central frequency that $\hbar\omega_{1}^{0}=\hbar\omega_{2}^{0}=2.06\text{ eV}$.}
\label{fig:Abs}
\end{figure}

\begin{figure*}[t]
\centering
\includegraphics[width=\textwidth]{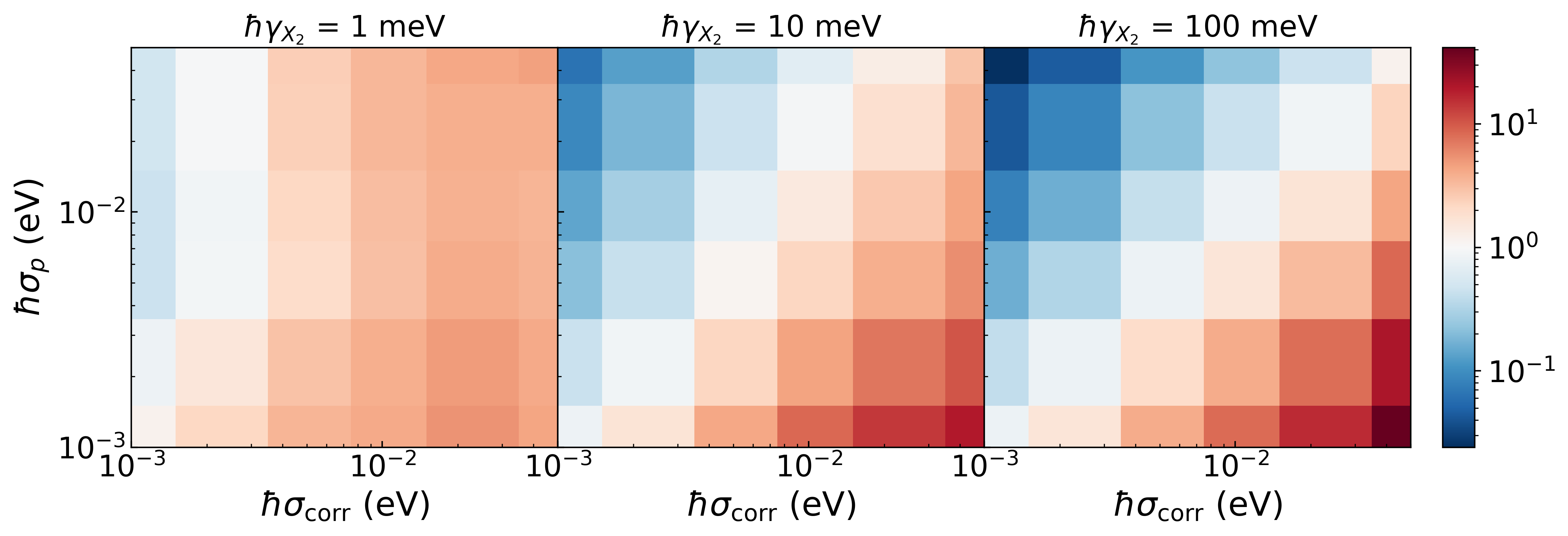}
\caption{Diagram of the efficiency ratio as a function of correlation and pump bandwidth of energy-time entangled quantum light for a multilevel CdSe/CdS core-shell QD.}
\label{fig:MLS}
\end{figure*}

\section{Results and Discussion}

We investigate biexciton generation in a quantum dot initially in its ground state using time-dependent perturbation theory. The single-exciton and biexciton responses are evaluated to second and fourth order in the light--matter interaction, respectively, retaining excitation through real intermediate excitonic states and interference among alternative two-photon excitation pathways. This framework allows us to determine how the pump bandwidth, photon correlation bandwidth, and excitonic dephasing control biexciton selectivity. The complete Hamiltonian and nonlinear-response derivation are provided in the Methods section.

We first introduce the light-field parameters that govern the excitation dynamics. For quantum light excitation, we consider entangled photon pairs generated via SPDC, described by the joint spectral amplitude $\Phi(\omega_1,\omega_2)$, which characterizes the temporal and energetic correlations between the photons:\cite{raymer2021entangled}
\begin{equation}
\begin{aligned}
\Phi(\omega_1,\omega_2) = {} & \mathcal{N}^{\text{QM}} \exp\!\left[-(\omega_1+\omega_2-\omega_1^0-\omega_2^0)^2 T^2 \right] \\
& \times \mathrm{sinc}\!\left[(\omega_1-\omega_2-\omega_1^0+\omega_2^0)\tau \right],
\label{eq:QL_JSA}
\end{aligned}
\end{equation}
where $T$ is the pump pulse duration and $\tau$ is the entanglement time. In the frequency domain, the pump bandwidth is $\sigma_p = 2\pi/T$, and the correlation bandwidth is $\sigma_{\text{corr}} = 2\pi/\tau$. 
In practical implementations, the correlation bandwidth $\sigma_{\text{corr}}$ can be flexibly tuned primarily by engineering the phase-matching conditions of the non-linear source,\cite{schlawin2018entangled, dorfman2021hong} or extrinsically through post-processing spectral filtering. This provides access to a wide experimental range from less than $1\text{ meV}$ up to several hundred $\text{meV}$~\cite{javid2021ultrabroadband,nasr2008ultrabroadband,steinlechner2012high}, a window that fully encompasses the parameter regime investigated in our simulations.
A longer pump duration $T$ leads to a narrower pump bandwidth, imposing a stronger constraint on the total photon energy, while a shorter entanglement time $\tau$ yields a broader correlation bandwidth, corresponding to tighter temporal correlations between the photons.\cite{raymer2021entangled} The parameters $\omega_1^0$ and $\omega_2^0$ denote the central frequencies, and $\mathcal{N}^{\text{QM}}$ is a normalization constant ensuring $\int\!\!\int d\omega_1 d\omega_2 \, |\Phi(\omega_1,\omega_2)|^2 = 1$.

To assess the role of quantum correlations, we also consider a classical coherent pulse with spectral envelope:
\begin{equation}
\psi(\omega) = \mathcal{N}^{\text{CM}} [ e^{-(\omega-\omega_1^0)^2 T^2} + e^{-(\omega-\omega_2^0)^2 T^2}],
\label{eq:CL_JSA}
\end{equation}
which exhibits the same power spectrum as biphoton state in the separable limit when $\Phi(\omega_1,\omega_2)=\psi(\omega_1)\psi(\omega_2)$.
As in the quantum case, $\mathcal{N}^{\text{CM}}$ ensures that $\int d\omega \, |\psi(\omega)|^2 = 1$. This classical field reproduces the same pump bandwidth as the quantum case but lacks the intrinsic temporal correlations of photons.

In Fig.~\ref{fig:light_profile}, we show the frequency-domain profiles of the photon field under the degenerate condition $\omega_1^0 = \omega_2^0$. Panel (a) shows the absolute joint spectral intensity $|\Phi(\omega_1,\omega_2)|$ of the entangled photons, which exhibits a characteristic anti-diagonal structure. In contrast, panel (b) displays the cross-product of the classical absolute spectral profiles $\left|\psi\left(\omega_{1}\right)\psi\left(\omega_{2}\right)\right|$, which illustrates a symmetric, uncorrelated distribution. The difference between features reflects energy conservation in the SPDC process, where the sum frequency $(\omega_1+\omega_2)$ is constrained by the pump bandwidth, while the frequency difference $(\omega_1-\omega_2)$ is governed by phase-matching conditions, encoded in the entanglement time $\tau$.

To this end, we begin with a minimal three-level model consisting of a ground state, a single bright excitonic state, and a biexcitonic state. 
Because the low-energy exciton and biexciton manifolds of strongly confined quantum dots are often more spectrally sparse than the corresponding higher-energy manifolds, the three-level system also provides a useful minimal model for understanding excitation of the lowest-energy biexciton when one bright intermediate exciton dominates the transition. The relevant energy scales and transition dipole moments are extracted from the excitonic and biexcitonic structure of the QD system described in Methods Section. Specifically, we consider a system with excitonic energy $E_S = 1.902\,\text{eV}$, biexcitonic energy $E_B = 3.801\,\text{eV}$. The transition dipole moments between the ground and excitonic states and between the excitonic and biexcitonic states are taken to be $\mu_{BS} = \mu_{Sg} = 3.476$ in atomic units (1 Debye $\approx$ 0.39 atomic units).  We vary the excitonic-state dephasing energy over $\hbar\gamma_S=0.5,1,5,10,50, \text{and }100\text{ meV}$, set the biexcitonic-state dephasing rate to $\gamma_B=2\gamma_S$, and the photon central frequencies to $\hbar \omega_1^0 = \hbar \omega_2^0 = E_B/2$. 
This simplified model allows us to isolate the essential physics governing correlated two-photon absorption and to systematically explore the parameter regimes in which quantum light provides a distinct advantage over classical excitation.

\begin{figure*}[t]
\centering
\includegraphics[width=\textwidth]{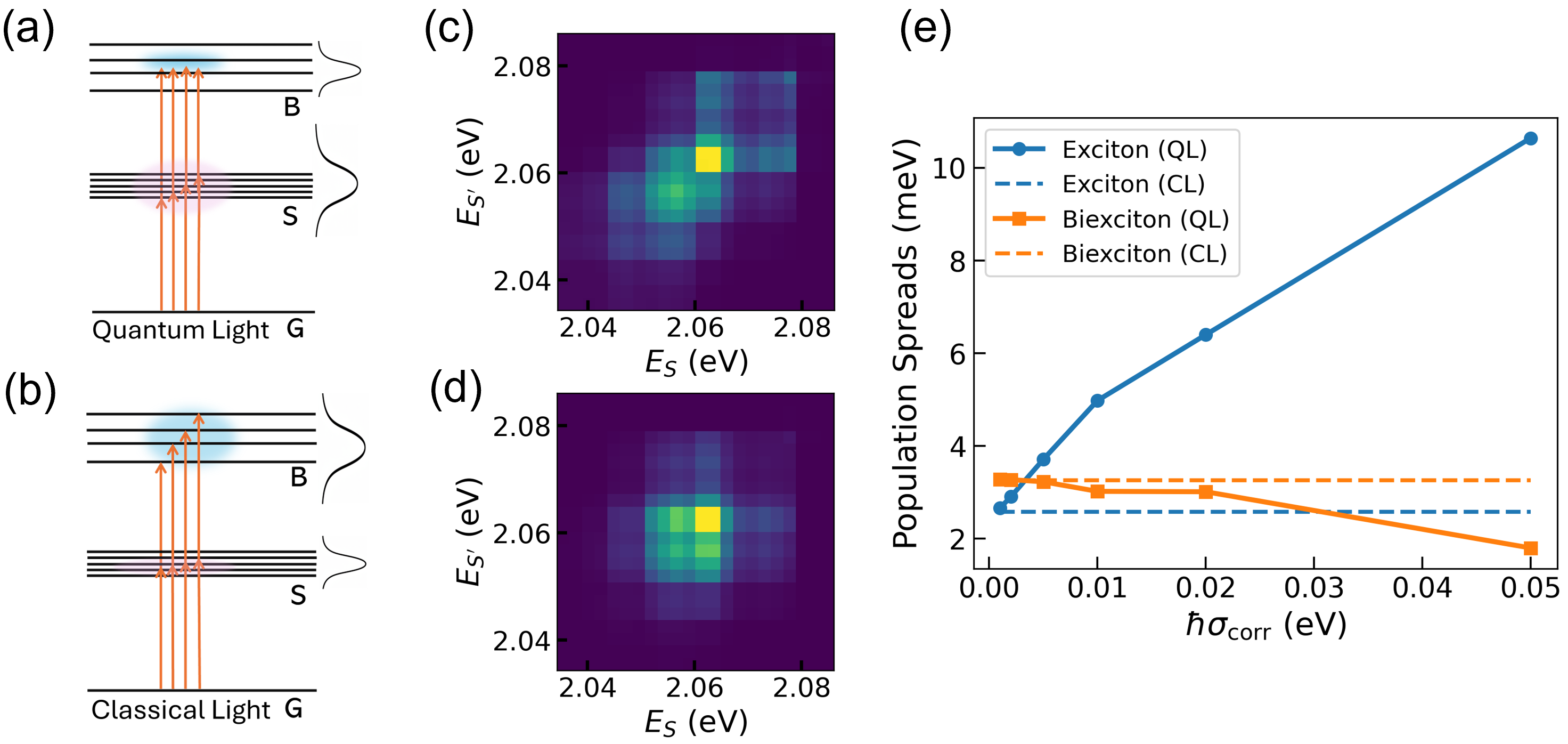}
\caption{Comparison of excitation ranges under classical and quantum light. (a,b) Schematic illustration of excitation pathways for quantum and classical light, respectively. The broad distribution of individual photon energies under quantum light provides access to a wider range of intermediate excitonic states, while the narrow sum-frequency distribution constrains the final biexciton energy. (c,d) Magnitudes of the single-exciton density-matrix elements, $|\rho_{SS'}|$, under quantum-light (QL) and classical-light (CL) excitation, respectively, calculated using $\hbar\gamma_S=0.001$~eV, $\hbar\sigma_p=0.01$~eV, and $\hbar\sigma_{\mathrm{corr}}=0.05$~eV. The diagonal elements, $\rho_{SS}$, represent exciton populations, whereas the off-diagonal elements, $\rho_{SS'}$ with $S\neq S'$, represent coherences between excitonic states. (e) Root-mean-square energy deviations $\sigma_S$ and $\sigma_B$ of the exciton and biexciton populations, respectively, as functions of the correlation bandwidth $\sigma_{\mathrm{corr}}$, calculated using $\hbar\gamma_S=0.001$~eV and $\hbar\sigma_p=0.001$~eV. QL and CL denote quantum-light and classical-light excitation, respectively. Increasing the correlation bandwidth under quantum-light excitation broadens the range of intermediate excitonic states while narrowing the distribution of final biexcitonic states.}
\label{fig:MLS_range}
\end{figure*}

In particular, we focus on the regime where the excitation is near-resonant with the singly excited state. Notably, this pathway directly mirrors the sequence exploited in multi-exciton cascade emission for generating entangled photon pairs. This regime is especially relevant for typical core-shell QDs, where the biexciton binding energy is comparable to or smaller than the excitonic dephasing width, leading to an intrinsic competition between resonant single-exciton absorption and correlated two-photon excitation pathways.

The performance of quantum light is quantified by comparing its ability to generate biexcitonic states relative to excitonic states with that of classical light. This describes how non-Poissonian the quantum light excitation is, independent of excitation power. Specifically, we define the efficiency ratio as
\begin{equation}
\eta = {\left[\frac{\sum_{B} P_{B}}{\sum_{S} P_{S}}\right]_{\mathrm{QL}}} {\Bigg /} {\left[\frac{\sum_{B} P_{B}}{\sum_{S} P_{S}}\right]_{\mathrm{CL}}},
\end{equation}
where $P_S$ and $P_B$ denote the populations of excitonic state $|S\rangle$ and biexcitonic state $|B\rangle$, respectively. The sums run over all excitonic and biexcitonic states included in the calculation, while the subscripts $\mathrm{QL}$ and $\mathrm{CL}$ denote excitation by quantum light and classical light, respectively. Thus, $\eta$ compares the biexciton-to-exciton population ratio generated by quantum light with the corresponding ratio generated by classical light. It is dimensionless and independent of the light intensity. A value of $\eta>1$ indicates that quantum light preferentially enhances biexciton generation relative to single-exciton excitation compared with classical light. Our goal is to determine whether quantum light can enhance the biexciton generation while suppressing the population of single excitons.

To systematically characterize the behavior of the quantum field, we construct a diagram in terms of the dimensionless correlation bandwidth and pump bandwidth, shown in  Fig.~\ref{fig:3LS_analysis}(a).  This allows us to identify the optimal regimes for quantum-enhanced biexciton generation. The bottom-left corner of the diagram corresponds to the regime where quantum light effectively reduces to the classical limit where $\eta\rightarrow 1$. A region of enhanced selective biexciton generation is highlighted by the black circle.

Fig.~\ref{fig:3LS_analysis}(b) shows the dependence of $\eta$ on the pump bandwidth for several correlation bandwidths, represented by lines of different colors and linestyles. We observe a nonmonotonic behavior of $\eta$, with a minimum at approximately $\sigma_p=5\gamma_S$, indicated by the dashed vertical line, which is largely independent of the correlation bandwidth. Classical light is constrained by the Fourier relation between temporal and spectral widths: a narrow pump bandwidth leads to low photon density in the time domain, whereas a broad pump bandwidth increases temporal photon density but reduces the effective excitation due to off-resonant contributions. As a result, classical light performs optimally only at intermediate pump bandwidths. By contrast, quantum light, with a tunable correlation bandwidth that is independent of the pump bandwidth, overcomes these limitations and maintains high performance away from the minimum in $\eta$, for both small and large values of $\sigma_p$.

Fig.~\ref{fig:3LS_analysis}(c) shows the efficiency ratio as a function of the correlation bandwidth. The efficiency $\eta$ initially increases with the correlation bandwidth, scaling approximately linearly with $\sigma_{\mathrm{corr}}/\gamma_S$, and saturates once the correlation bandwidth becomes sufficiently large. This saturation behavior indicates that when the photon arrival-time difference is much shorter than the system response time, no further enhancement in biexciton absorption can be achieved. 

When $\sigma_p$ is large, we find a decrease in $\eta$ before saturation. This behavior arises from the competition between two effects as the correlation bandwidth increases: (i) reduced photon arrival-time separation, which enhances biexciton generation, and (ii) lower on-resonant contribution at off-resonant excitation, which reduces the efficiency. As a result, an intermediate correlation bandwidth becomes optimal, especially when the excitonic dephasing rate $\gamma_S$ is small.

\begin{figure*}[t]
\centering
\includegraphics[width=\textwidth]{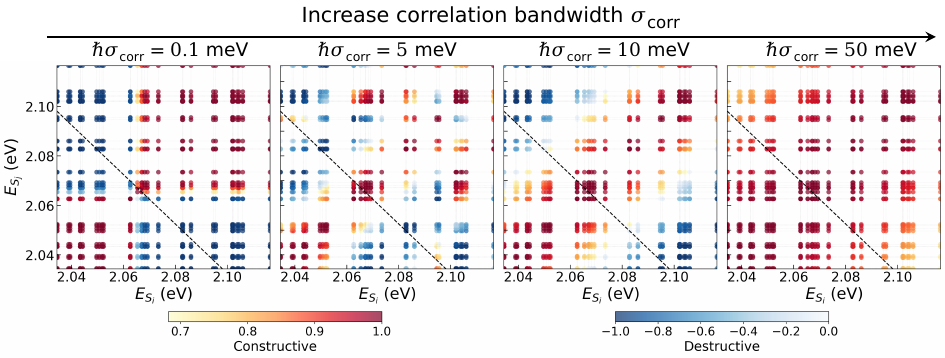}
\caption{Interference map between exciton pairs $S_{i}$ and $S_{j}$. Red (blue) indicates constructive (destructive) interference between excitation pathways contributing to biexciton formation. The dashed line corresponds to the energy conservation condition $E_{S_{i}}+E_{S_{j}}-E_{b}=\hbar\omega_{p}$. From left to right, the correlation bandwidth increases with values of $0.1, 5, 10$, and $50$~meV. As the bandwidth grows, a larger number of exciton pairs along the resonant energy line exhibit constructive interference, enhancing the biexciton excitation probability. \label{fig:interference}}
\end{figure*}

We next examine the performance of entangled quantum light in a more realistic, multilevel system, by considering a $3.9$ nm CdSe core with a $2$-monolayer (ML) CdS shell nanocrystal. The calculated classical absorption spectrum of the QD is shown in Fig.~\ref{fig:Abs}. The absorption onset is characterized by several distinct bright excitonic transitions, indicated by the vertical lines corresponding to their oscillator strengths.\cite{brosseau2023ultrafast,ghosh2025atomistically} The photon central energies were chosen to match the $X_2$ transition energy (indicated by the yellow arrow), such that $\hbar \omega_1^0 = \hbar \omega_2^0 = 2.060\,\text{eV}$.

In the simulation, we include $30$ excitonic states, resulting in $465$ biexcitonic states within the relevant excitation window, including configurations formed from two excitons with the same spatial wavefunction but opposite spin configurations.  The parameterization of the multilevel Hamiltonian is provided in the Supporting Information. The efficiency ratio of quantum light for this multilevel system is presented in Fig.~\ref{fig:MLS} for different dephasing rates $\gamma_{X_2}$ using excitonic state $X_2$ as the reference state. The dephasing rates of the remaining excitonic states are scaled according to their calculated reorganization energies, $\gamma_S=\gamma_{X_2}\lambda_S/\lambda_{X_2}$, as described in the Supporting Information. The overall dependence of $\eta$ closely resembles that of the three-level system shown in Fig.~\ref{fig:3LS_analysis}, although here the axes are expressed in physical units and span a narrower range of energies. In general, increasing the correlation bandwidth significantly enhances the performance of quantum light. In contrast, classical light performs well only when $\sigma_p$ is very large, which is inevitably accompanied by a loss of spectral selectivity. 

To elucidate the underlying mechanism, it is instructive to examine the role of strong temporal correlations, specifically how entanglement modifies the excitation dynamics and mediates interference among multiple excitation pathways. We first examine the excitation energy range induced by classical and quantum light. We find that the range of excitonic states accessible under quantum light can be significantly broader than under classical light when the correlation bandwidth exceeds the pump bandwidth ($\sigma_{\mathrm{corr}} > \sigma_p$). In contrast, the excitation range of biexcitonic states can be considerably narrower than in the classical case, even though both share the same pump bandwidth. This seemingly counterintuitive behavior arises because entangled photons allow for a wide distribution of individual photon energies while still enforcing a strict constraint on their sum frequency. As a result, single-photon (exciton) transitions can access a broader range of energies, whereas two-photon (biexciton) transitions remain sharply confined by energy conservation, leading to enhanced spectral selectivity. In addition, the large correlation bandwidth conceals the which-path information of the excitation sequence. The resulting indistinguishability leads to an overall constructive interference among the multiple excitation pathways, a mechanism we will explore in detail below.
 A schematic illustration of this behavior is shown in Fig.~\ref{fig:MLS_range}(a) for quantum light and in Fig.~\ref{fig:MLS_range}(b) for classical light.  

Next, we analyze the excitation of single excitons in greater detail by computing the density matrix in the basis of singly excited states, $\rho_{SS'}$, using Eq.~\eqref{eq:single-TDPT} in Methods Section below. The resulting density matrices are shown in Fig.~\ref{fig:MLS_range}(c) and (d) for quantum and classical light, respectively, under the conditions $\hbar \gamma_{S}=0.001\,\text{eV}$, $\hbar \sigma_{p}=0.01\,\text{eV}$, and $\hbar \sigma_{\mathrm{corr}}=0.05\,\text{eV}$. As observed, the density matrix under classical light is more centered and exhibits a compact, symmetric structure, whereas under quantum light it is elongated along the anti-diagonal. 
This difference can be understood as follows. The diagonal elements represent populations, governed by the single-photon energy distribution. Since $\sigma_{\mathrm{corr}} > \sigma_p$, the accessible population range is broadened for quantum light. The off-diagonal elements ($S \neq S'$) correspond to coherences, which are controlled by the pump bandwidth. 
For classical light, the two-photon spectral distribution factorizes as $\Phi(\omega_1,\omega_2)=\psi(\omega_1)\psi(\omega_2)$, implying that the sum and difference frequencies share the same bandwidth. Consequently, the range of populations and coherences is identical, leading to the compact, symmetric structure. 

To further quantify the excitation profile, we define the population spreads for excitons and biexcitons as

\begin{equation} \begin{split} \sigma_{S} &= \sqrt{ \frac{\sum_{S} P_{S} \left(E_{S}-\hbar\omega_{1}^{0}\right)^{2}} {\sum_S P_S} }, \\ \sigma_{B} &= \sqrt{ \frac{\sum_{B} P_{B} \left(E_{B}-\hbar\omega_{p}\right)^{2}} {\sum_B P_B} }. \end{split} \label{eq:population_spreads} \end{equation}
Here, $P_S$ and $P_B$ are the populations of excitonic state $|S\rangle$ and biexcitonic state $|B\rangle$, while the denominators normalize the corresponding population distributions. $E_S$ and $E_B$ are the energies of excitonic state $|S\rangle$ and biexcitonic state $|B\rangle$, respectively. Under the degenerate condition considered here, $\omega_1^0=\omega_2^0=\omega_p/2$. Accordingly, $\sigma_S$ and $\sigma_B$ measure the root-mean-square detuning of the populations from the excitation energies $\omega_{1}^{0}$ for single-photon absorption and $\omega_{p}$ for two-photon absorption, respectively. The results are shown in Fig.~\ref{fig:MLS_range}(e), where “QL” and “CL” denote excitation by quantum light and classical light, respectively. The calculations are performed with $\hbar \gamma_{S}=0.001\,\text{eV}$ and $\hbar \sigma_{p}=0.001\,\text{eV}$, chosen to clearly reveal the dependence on the correlation bandwidth. We find that under quantum light excitation, increasing the correlation bandwidth leads to an increase in $\sigma_{S}$, while $\sigma_{B}$ decreases, indicating a more energy-selective biexciton excitation. This highlights the distinct role of quantum correlations in shaping excitation pathways.

To further elucidate the role of photon correlations, we analyze biexciton population given by Eq.~\eqref{eq:PB} in Methods Section below in the limit of a narrow pump bandwidth, $\sigma_p \to 0$, where the integral can be evaluated analytically. For the case $\omega_1^0 = \omega_2^0$, we obtain

\begin{equation}
\begin{aligned}
P_B &= \left| \mathcal{N}_t^{\mathrm{QM}} \frac{\sqrt{\omega_2^0\omega_1^0}} {\hbar\varepsilon_0 c A_0} \frac{\gamma_B/\pi} {\left(\omega_{Bg}-\omega_2^0-\omega_1^0\right)^2+\gamma_B^2} \right. \\ &\qquad \left. \times \sum_S \mu_{BS}\mu_{Sg} \left[ \frac{ e^{-i\left(\omega_{Sg}-\omega_1^0-i\gamma_S\right)\tau}-1 }{ \omega_{Sg}-\omega_1^0-i\gamma_S } \right] \right|^2, 
\label{eq:PB_limit}
\end{aligned} 
\end{equation}
where $\mathcal{N}_t^{\mathrm{QM}}$ is the normalization factor of the biphoton joint temporal amplitude, as defined in the Supporting Information. The quantities $\omega_{Sg}$ and $\omega_{Bg}$ are the exciton and biexciton transition frequencies, respectively, while $\mu_{Sg}$ and $\mu_{BS}$ are the corresponding transition dipole moments, as defined in the Methods Section. The sum over $S$ includes all intermediate excitonic states contributing to the formation of biexciton $B$. Finally, $A_0$ is the effective beam area. $\epsilon_0$ is the vacuum permittivity, $c$ is the speed of light.
The classical light limit is recovered by taking the entanglement time $\tau = 1/\sigma_{\mathrm{corr}} \to \infty$, in which case the phase factor simplifies to
\begin{equation}
\frac{e^{-i\left(\omega_{Sg}-\omega_1^0-i\gamma_S\right)\tau}-1}{\omega_{Sg}-\omega_1^0-i\gamma_S}
\;\longrightarrow\;
\frac{-1}{\omega_{Sg}-\omega_1^0-i\gamma_S}.
\end{equation}
More generally, this term can be expressed in polar form as
\begin{equation}
\frac{e^{-i\left(\omega_{Sg}-\omega_1^0-i\gamma_S\right)\tau}-1}{\omega_{Sg}-\omega_1^0-i\gamma_S}
= |T_S| e^{i\phi_S},
\end{equation}
where $|T_S|$ is the magnitude and $\phi_S$ is the phase associated with the intermediate excitonic state $S$.

We now consider a biexcitonic state $\lvert B \rangle$ formed from two excitonic states $\lvert S_i \rangle$ and $\lvert S_j \rangle$. The excitation can proceed via two pathways: $\lvert g \rangle \rightarrow \lvert S_i \rangle \rightarrow \lvert B \rangle$ or $\lvert g \rangle \rightarrow \lvert S_j \rangle \rightarrow \lvert B \rangle$. The total contribution from these pathways depends on their relative phase difference, which we quantify through the interference factor
\begin{equation}
M_{ij} = \cos\left(\phi_{S_i} - \phi_{S_j}\right).
\end{equation}
When $M_{ij} > 0$, the interference is constructive, with values approaching $1$ indicating stronger constructive interference. Conversely, $M_{ij} < 0$ corresponds to destructive interference, with values approaching $-1$ indicating stronger suppression.

Fig.~\ref{fig:interference} shows the interference map, where the axes correspond to the exciton energies for states $\lvert S_i \rangle$ and $\lvert S_j \rangle$. The dashed line corresponds to the condition $E_{S_i} + E_{S_j} - E_b = \hbar \omega_p$. Exciton pairs lying near this line satisfy energy conservation and therefore contribute efficiently to biexciton formation. From left to right panels, the correlation bandwidth $\sigma_{\mathrm{corr}}$ increases (equivalently, the entanglement time $\tau$ decreases), with values of $0.1$, $5$, $10$, and $50\,\text{meV}$, corresponding to $\tau = 6583$, $132$, $65.8$, and $13.2\,\text{fs}$, respectively. The value of $\tau$ in the left panel corresponds to the classical light limit, where $\sigma_{\mathrm{corr}} \to 0$ ($\tau \to \infty$).

\begin{figure}[t]
\centering
\includegraphics[width=8cm]{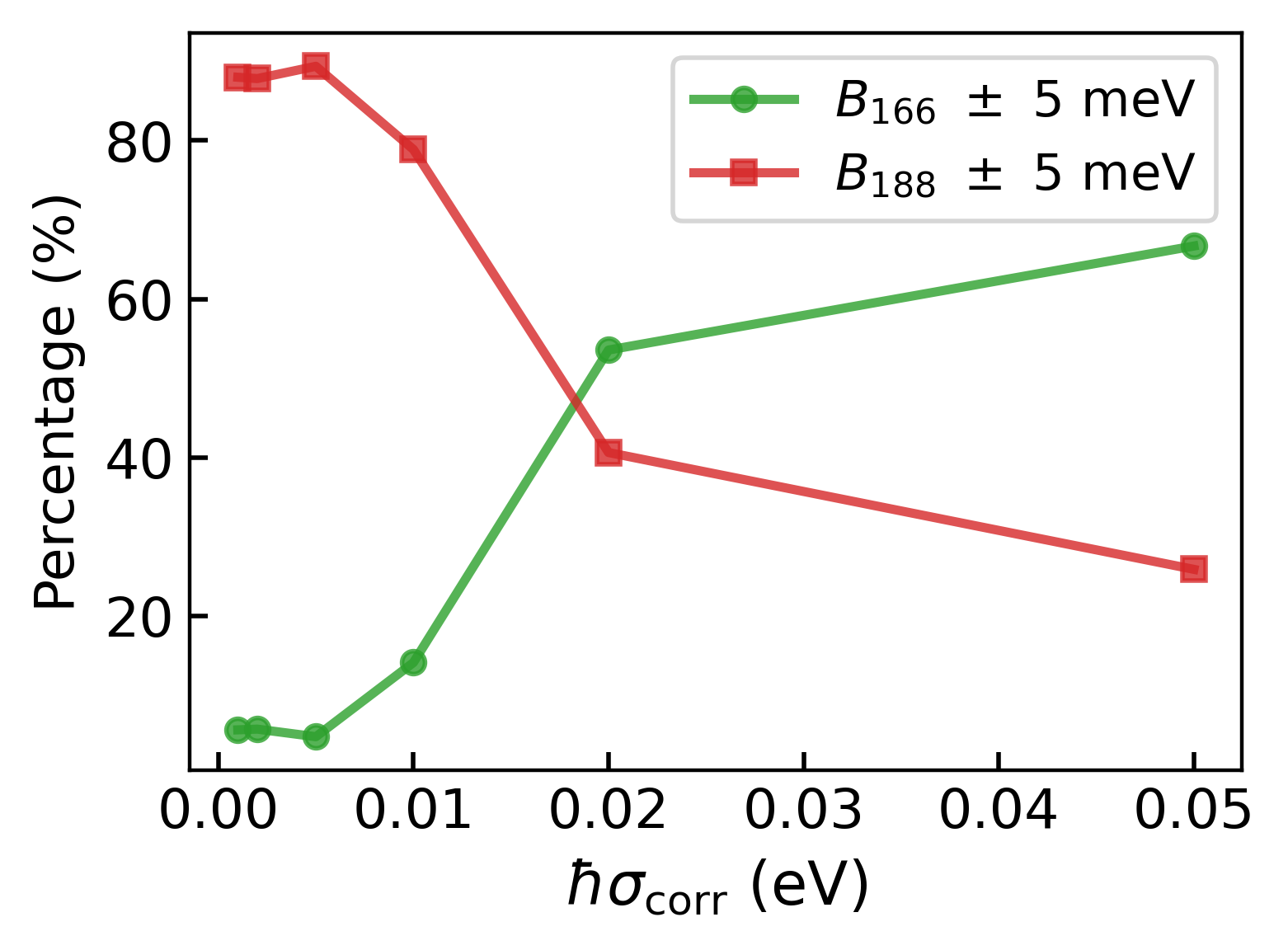}
\caption{Population percentages of biexciton states $B_{166}$ ($E_{B_{166}}=4.11eV$) and $B_{188}$ ($E_{B_{188}}=4.13eV$) with a $5\text{meV}$ energy window, plotted as a function of correlation width $\sigma_\mathrm{corr}$ at fixed pump bandwidth $\hbar\sigma_{p}=0.01\text{eV}$.}
\label{fig:selective}
\end{figure}

We find that along the energy-conserving line, the number of constructively interfering exciton pairs increases with increasing $\sigma_{\mathrm{corr}}$. This indicates that larger correlation bandwidths not only increase the number of accessible excitation pathways within the excitonic manifold, but also enhance the fraction of constructively interfering pathways, thereby leading to an increased biexciton population. At small correlation bandwidths, biexciton generation is limited by the scarcity of constructively interfering pathways. As the correlation bandwidth increases, a larger number of excitation pathways become accessible due to relaxed temporal constraints, while the strict constraint on the sum frequency enforces energy conservation. As a result, transitions that satisfy the biexciton resonance condition are preferentially enhanced. This behavior motivates us to explore whether selective excitation of specific biexcitonic states can be achieved by tuning the correlation bandwidth.

In Fig.~\ref{fig:selective}, we plot the population fractions of two biexciton states, $\lvert B_{166} \rangle$ ($E_{B_{166}} = 4.11\,\text{eV}$) and $\lvert B_{188} \rangle$ ($E_{B_{188}} = 4.13\,\text{eV}$), within an energy window of $5\,\text{meV}$, under the conditions $\hbar \sigma_p = 0.01\,\text{eV}$ and $\hbar \gamma_S = 0.001\,\text{eV}$, as a function of the correlation bandwidth $\sigma_{\mathrm{corr}}$. The excitation pathway leading to $\lvert B_{166} \rangle$ is initially suppressed due to destructive interference at small correlation bandwidths, despite satisfying energy conservation more closely. In contrast, $\lvert B_{188} \rangle$ benefits from constructive interference but deviates more from the energy conservation condition. Both states are optically active, leading to competing population channels. As $\sigma_{\mathrm{corr}}$ increases (corresponding to shorter entanglement times), a clear population transfer from $\lvert B_{188} \rangle$ to $\lvert B_{166} \rangle$ is observed. This behavior demonstrates that tuning the correlation bandwidth provides a mechanism for selectively enhancing specific biexcitonic states, offering a potential tool for probing and controlling the fine structure of quantum dot systems.

More broadly, the mechanism identified in this paper is not limited to the specific SPDC source considered here. Quantum-light sources with appropriately shaped, nonfactorizable two-photon joint spectral amplitudes may provide similar control over multiphoton excitation pathways, although the optimal conditions will depend on source-specific spectral properties, purity, and photon distinguishability. 
Besides, non-resonant excitation may additionally involve virtual-state contributions, phonon-assisted transitions, and population relaxation, which warrant further investigation.
Likewise, while the phenomenological dephasing rates capture the net loss of optical coherence, processes such as population relaxation, spectral diffusion, trapping, radiative decay, and Auger recombination may require an explicit treatment. Thus, the qualitative role of photon correlations is expected to be general, whereas quantitative predictions remain dependent on the light source, excitation regime, and material-specific relaxation mechanisms.

\section{Conclusions}
In this study, we investigated the role of energy-time entangled quantum light in enhancing and controlling biexciton generation in semiconductor QDs. By introducing a dimensionless efficiency parameter that compares biexciton-to-exciton generation under quantum and classical light, we established a clear and intensity-independent metric to evaluate quantum advantages. Using a simplified three-level model, we showed that increasing the correlation bandwidth does not necessarily improve performance. Instead, the performance ratio typically saturates and, at large values of $\sigma_{p}/\gamma_{S}$, exhibits a trade-off between enhanced temporal overlap and off-resonant excitation, leading to an optimal intermediate regime for biexciton generation.

To understand how structural complexity influences this dynamic, we compared the three-level model to a many-level system characterized by a broad spread of intermediate excitonic states. Using a realistic CdSe/CdS core-shell QD as a representative example, we demonstrated that the enhanced biexciton generation stems fundamentally from this multiplicity of intermediate states, rather than from material-specific properties of the QD itself. Furthermore, by comparing excitation under quantum and classical light, we showed that increasing the correlation bandwidth in the quantum case leverages this spread of intermediate states to open a larger manifold of excitation pathways. The resulting enhanced constructive interference among these pathways leads to a more efficient population of biexcitonic states relative to classical excitation. Moreover, the interplay between constructive interference and energy conservation enables selective excitation of closely spaced biexciton states, providing a route to probe fine structure and control multi-excitation selectivity. More broadly, the ability to independently tune temporal correlations and spectral properties highlights the potential of entangled quantum light for selective state preparation and quantum control in complex multilevel systems.

\section{Methods}
\subsection{Model Hamiltonian}
We begin by defining the total Hamiltonian of the quantum dot (QD), the light field, and their coupling as:
\begin{equation}
H=H_\text{QD}+H_\text{field}+V,
\end{equation}
where the QD is described by $H_\text{QD}$ and is coupled to the photon field $H_\text{field}$ via the interaction $V$. $H_\text{QD}$ is characterized by a ground state $\lvert g \rangle$ with energy $E_g = 0$, a manifold of excitonic states $\lvert S \rangle$ with energies $E_S$, and a manifold of biexcitonic states $\lvert B \rangle$ with energies $E_B$: 
\begin{equation}
H_{\text{QD}}=E_{g}\left|g\right\rangle \left\langle g\right|+\sum_{S}E_{S}\left|S\right\rangle \left\langle S\right|+\sum_{B}E_{B}\left|B\right\rangle \left\langle B\right|.
\end{equation}
For simplicity, we neglect the couplings between the ground and the doubly excited manifold of states and between singly and doubly excited states, which give rise to multiexciton generation and Auger recombination.  The former has a negligible probability at the onset energies~\cite{rabani2010theory} and the latter occurs on timescales longer than those relevant in the absorption discussed here~\cite{philbin2018electron}.

The excitonic energies $E_{S}$ were parameterized using the semi-empirical pseudopotential method~\cite{jasrasaria2022simulations} combined with the Bethe-Salpeter equation (BSE)~\cite{rohlfing2000electron}, within a static screening approximation~\cite{eshet2013electronic}. The corresponding exciton wave functions are given in terms of combinations of electron-hole pair states $\left|S\right\rangle =\sum_{ai}c_{ai}^{S}a_{a}^{\dagger}a_{i}\left|g\right\rangle \otimes\left|\chi_{S}\right\rangle$,  where $c_{ai}^{S}$ are the BSE coefficients and $\left|\chi_{S}\right\rangle =\frac{1}{\sqrt{2}}\left[\left|\uparrow_a\uparrow_i\right\rangle +\left|\downarrow_a\downarrow_i\right\rangle \right]$ represents the spin singlet configuration. Biexcitonic states were modeled as two spatially non-interacting but spin correlated excitons, $\left|B\right\rangle =\sum_{bj}\sum_{ai}c_{bj}^{S_{1}}c_{ai}^{S_{2}}a_{b}^{\dagger}a_{j}a_{a}^{\dagger}a_{i}\left|g\right\rangle \otimes\left|\chi_{B}\right\rangle $, where $\left|\chi_{B}\right\rangle =\frac{1}{\sqrt{2}}\left[\left|\uparrow_b\uparrow_j,\downarrow_a\downarrow_i\right\rangle +\left|\downarrow_b\downarrow_j,\uparrow_a\uparrow_i\right\rangle \right]$ represents the spin singlet configuration~\cite{philbin2018electron}. This is justified by the observation that electron-hole correlations within individual excitons significantly outweigh those between excitons, particularly in the strong confinement regime~\cite{amgar2025correlations,moody2013influence}. For the $3.9$~nm diameter CdSe/$2$~ML CdS core-shell QD studied here, we adopt a constant, experimentally derived biexciton binding energy of $E_b \approx 3\text{ meV}$~\cite{scharf2026unraveling}, to account for inter-excitonic Coulomb and exchange interactions generally. Such that a biexciton composed of excitonic states $\left|S_1\right\rangle$ and $\left|S_2\right\rangle$ has an energy of $E_B = E_{S_1} + E_{S_2} - E_b$. Minor state-dependent variations in $E_b$ are small relative to the biphoton correlation bandwidth and do not qualitatively alter the predicted interference patterns.

The quantized electromagnetic field is expressed as a continuum of frequency modes:
\begin{equation}
H_{\text{field}}=\sum_{i}\int d\omega\hbar\omega a_{i}^{\dagger}\left(\omega\right)a_{i}\left(\omega\right),
\end{equation}
where $a_{i}^{\dagger}\left(\omega\right)$ and $a_{i}\left(\omega\right)$
are the creation and annihilation operators for mode $i$ with frequency
component $\omega$, satisfying the commutation relationship $\left[a_{i}\left(\omega\right),a_{j}^{\dagger}\left(\omega'\right)\right]=\delta_{ij}\delta\left(\omega-\omega'\right)$. 

The interaction between the field and the QD is described within the dipole and rotating wave approximations (RWA), such that the interaction Hamiltonian is given by (operators are expressed in the interaction picture unless stated otherwise):
\begin{equation}
V_I(t) = \hat{d}(t)\cdot\hat{E}(t) \approx \hat{E}^{(+)}(t) \hat{d}^{(-)}(t) + \text{h.c.},
\label{eq:V_I}
\end{equation}
where the field operator is decomposed as $\hat{E}(t) = \hat{E}^{(+)}(t) + \hat{E}^{(-)}(t)$, with $\hat{E}^{(-)}(t) = [\hat{E}^{(+)}(t)]^{\dagger}$. The positive-frequency component of the electric field operator is given by
\begin{equation}
\hat{E}^{(+)}(t) = \sum_i \int d\omega \, \sqrt{\frac{\hbar \omega}{4\pi \epsilon_0 c A_0}} \, \hat{a}_i(\omega) e^{-i\omega t},
\end{equation}
where $i$ labels the photon mode, $\epsilon_0$ is the vacuum permittivity, $c$ is the speed of light, and $A_0$ is the effective beam area~\cite{raymer2021entangled}.

Similarly, the dipole operator is written as $\hat{d}(t) = \hat{d}^{(+)}(t) + \hat{d}^{(-)}(t)$, with $\hat{d}^{(-)}(t) = [\hat{d}^{(+)}(t)]^{\dagger}$. The lowering part of the dipole operator is given by
\begin{equation}
\hat{d}^{(-)}(t) = \sum_S \mu_{Sg} e^{i\omega_{Sg} t} \lvert S \rangle \langle g \rvert
+ \sum_{B,S} \mu_{BS} e^{i\omega_{BS} t} \lvert B \rangle \langle S \rvert,
\end{equation}
where $\omega_{nm} = (E_n - E_m)/\hbar$, $\mu_{Sg} = \langle S \rvert \hat{d} \lvert g \rangle$, and $\mu_{BS} = \langle B \rvert \hat{d} \lvert S \rangle$ denote the transition dipole moments for the $\lvert g \rangle \to \lvert S \rangle$ and $\lvert S \rangle \to \lvert B \rangle$ transitions, respectively. Note that $\mu_{BS} \neq 0$ only when the excitonic state $S$ is a constituent of the biexcitonic state $B$.

 For quantum light excitation, the SPDC generated two-mode squeezed state of twin photons, labeled $1$ and $2$, is given by
\begin{equation}
\lvert \Psi \rangle = \int\!\!\int d\omega_1 d\omega_2 \, \Phi(\omega_1,\omega_2) \, a_1^\dagger(\omega_1) a_2^\dagger(\omega_2) \lvert 0 \rangle,
\end{equation}
where the joint spectral amplitude $\Phi(\omega_1,\omega_2)$ is given in Eq.~\eqref{eq:QL_JSA}. For a classical coherent pulse, the field operator is replaced by a classical field:
\begin{equation}
E(t) = \int d\omega \, \sqrt{\frac{\hbar \omega}{4\pi \varepsilon_0 c A_0}} \, \psi(\omega) e^{-i\omega t} + \text{h.c.},
\end{equation}
with spectral envelope given in Eq.~\eqref{eq:CL_JSA}.

The present model focuses on photon absorption and treats the incident field as a prescribed, undepleted input. It therefore neglects depletion and reshaping of the incident photon wavepacket, electromagnetic back-action on the photon field, stimulated re-emission, and repeated absorption--emission cycles. Environmental effects are incorporated phenomenologically through exciton and biexciton dephasing rates, as described below in the dynamics section to account for phonon couplings.

\subsection{Nonlinear Response and Exciton Dynamics}
To describe the generation of excitonic and biexcitonic states, we employ time-dependent perturbation theory.\cite{raymer2021entangled} Phonon-mediated dephasing as well as additional environmental loss or scattering channels are treated phenomenologically by assigning a dephasing rate $\gamma_{S/B}$ to excitonic and biexcitonic states. By varying the temperature, the dephasing rate spans a range from several hundred $\mu\text{eV}$ to several hundred $\text{meV}$.\cite{lin2023theory,peng2026photoluminescence} 

The density matrix in the interaction picture is given by
\begin{equation}
\rho_I(t) = U_I(t,t_0)\,\rho(t_0)\,U_I^\dagger(t,t_0),
\end{equation}
where the initial density matrix is $\rho(t_0) = \lvert g \rangle \langle g \rvert \otimes \lvert \Psi \rangle \langle \Psi \rvert$, and $U_I(t,t_0)$ is the time-evolution operator expressed via the Dyson series,
\begin{equation}
U_I(t,t_0) = \mathcal{T} \exp\left[-\frac{i}{\hbar}\int_{t_0}^{t} dt_1 \, V_I(t_1)\right],
\end{equation}
with $\mathcal{T}$ the time-ordering operator and $V_I(t) = \hat{d}(t)\cdot\hat{E}(t)$ the interaction Hamiltonian in the interaction picture (cf., Eq.~\eqref{eq:V_I}). The single-photon absorption coherence and population are obtained by tracing over the photon field degrees of freedom,
\begin{equation}
\rho_{I,SS'}(t) = \mathrm{Tr}_f \left[ \langle S \rvert \rho_I(t) \lvert S' \rangle \right].
\end{equation}
Expanding the evolution operator to second order, the excitonic density matrix elements $\rho_{I,SS'}(t)$ are given by
\begin{equation}
\begin{aligned}
\rho_{I,SS'}(t) 
&= \frac{1}{\hbar^{2}} \int_{-\infty}^{t} dt_{2} \int_{-\infty}^{t_{2}} dt_{1} \\
& \times \mathrm{Tr}_{f} \Big[
\langle S | V_{I}(t_{1}) \rho(t_{0}) V_{I}(t_{2})+\text{h.c.} | S' \rangle
\Big].
\label{eq:single-TDPT}
\end{aligned}
\end{equation}
In the steady-state limit $t \to \infty$ with $t_0 = -\infty$, the exciton population $P_S$ simplifies to
\begin{equation}
\begin{aligned}
P_{S} = & \frac{|\mu_{Sg}|^{2}}{\hbar \varepsilon_{0} c A_{0}}
\int\int d\omega_{1} d\omega_{2} \, |\Phi(\omega_{1},\omega_{2})|^{2} \\
& \times \left[
\frac{\omega_{1} \gamma_{S}}{\gamma_{S}^{2} + (\omega_{Sg} - \omega_{1})^{2}}
+ \frac{\omega_{2} \gamma_{S}}{\gamma_{S}^{2} + (\omega_{Sg} - \omega_{2})^{2}}
\right].
\label{eq:PS}
\end{aligned}
\end{equation}

The biexciton density matrix is evaluated by neglecting indirect absorption processes and retaining only second-order transitions mediated by the intermediate excitonic manifold. Using the same notation as for single-photon absorption, the biexcitonic density matrix elements are defined as
\begin{equation}
\rho_{I,BB'}(t) = \mathrm{Tr}_f \left[ \langle B \vert \rho_I(t) \vert B' \rangle \right].
\end{equation}
Expanding the evolution operator to fourth order, the biexcitonic density matrix elements are given by
 \begin{equation}
 \begin{aligned}
 \rho_{I,BB'}(t)&= \frac{1}{\hbar^{4}}
 \int_{-\infty}^{t}dt_{4}\int_{-\infty}^{t_{4}}dt_{3}\int_{-\infty}^{t_{3}}dt_{2}\int_{-\infty}^{t_{2}}dt_{1} \\
 & \times \mathrm{Tr}_f \Big[
 \langle B \vert V_I(t_{4}) V_I(t_{3}) \rho(t_0) V_I(t_{1}) V_I(t_{2}) \quad \\
& + V_I(t_{4}) V_I(t_{2}) \rho(t_0) V_I(t_{1}) V_I(t_{3}) \quad\\
 &  + V_I(t_{4}) V_I(t_{1}) \rho(t_0) V_I(t_{2}) V_I(t_{3})+\text{h.c.} \vert B' \rangle
 \Big].
\end{aligned}
\end{equation}
In the steady-state limit $t \to \infty$, we focus on the biexciton population $P_B = \rho_{I,BB}$, which becomes
\begin{equation}
\label{eq:PB}
\begin{aligned}
P_{B}
&=\left|
\frac{\sum_{S}\mu_{BS}\mu_{Sg}}{2\hbar\varepsilon_{0}cA_{0}}\int\int d\omega_{2} d\omega_{1} \right. \qquad\\
& \times\frac{\sqrt{\omega_{2}\omega_{1}}\,\Phi(\omega_{1},\omega_{2})\,\gamma_{B}/\pi}
{(\omega_{Bg}-\omega_{2}-\omega_{1})^{2}+\gamma_{B}^{2}} \qquad \\
&\left.\times\left[\frac{1}{\omega_{1}-\omega_{Sg}+i\gamma_{S}}
+\frac{1}{\omega_{2}-\omega_{Sg}+i\gamma_{S}}\right]\right|^{2}.
\end{aligned}
\end{equation}

For classical light excitation, $\Phi(\omega_1,\omega_2)$ is replaced by $\psi(\omega_1)\psi(\omega_2)$ in both $P_S$ and $P_B$, e.g., Eqs.~\eqref{eq:PS} and \eqref{eq:PB}. Since the electric field amplitude scales as $E \propto \sqrt{1/A_0}$, it follows that $P_S \propto E^{2}$ and $P_B \propto E^{4}$.

\section{Supporting Information}
Supporting Information. Quantum dot configurations and properties, bandwidth-dependent scaling in a three-level model system, bandwidth dependence in realistic multi-level quantum dots, derivation of exciton and biexciton dynamics, narrow pump bandwidth limit.

\section*{Associated Content}
A preprint version of this manuscript is available.~\cite{peng2026selective}

\begin{acknowledgement}
The National Science Foundation, Division of Chemistry, under the Chemical Theory, Models and Computational Methods (CTMC) program (Grant No.~CHE-2449564) supported the work on model development by E.R. and K.P. and the U.S. National Science Foundation Science and Technology Center (STC) for Integration of Modern Optoelectronic Materials on Demand (IMOD) under Cooperative Agreement No. DMR-2019444 supported the calculations on nanocrystals by K.P. The work was also supported by the U.S. Department of Energy, Office of Science, Office of Basic Energy Sciences, Materials Sciences and Engineering Division, under Contract No. DE-AC02-05CH11231 within the Fundamentals of Semiconductor Nanowire Program (KCPY23). This research used resources of the National Energy Research Scientific Computing Center (NERSC), a U.S. Department of Energy Office of Science User Facility supported under Contract No.~DE-AC02-05CH11231, through NERSC award BES-ERCAP0032503. E.R. acknowledges support from the Israel Science Foundation (Grant No.~4085/25). 
\end{acknowledgement}
 
\bibliography{reference}
\end{document}


\section{Quantum Dot Configurations and Properties}

\begin{figure}[t]
\includegraphics[width=16cm]{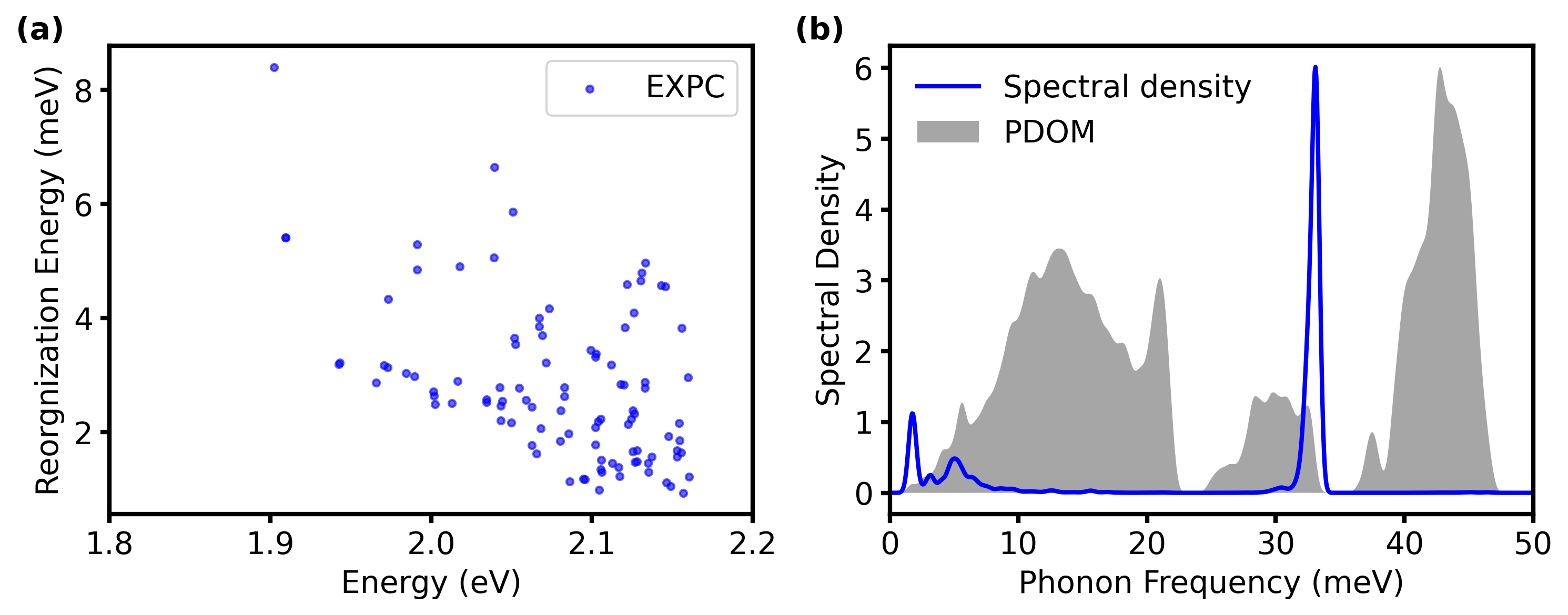}
\caption{(a) Reorganization energy $\lambda_{S}$ for excitonic states ranging from 1.9 to 2.2 eV for a $3.9$ nm CdSe core with a $2$ML CdS shell QD. Each point represents one excitonic state. (b) Phonon density of modes (PDOM) $\rho_{P}\left(\omega\right)$ and spectral density $J\left(\omega\right)$ for lowest bright excited state, a Gaussian broadening with a standard deviation of $0.3\text{ meV}$ is applied here. \label{fig:QD_property}}
\end{figure}

We consider a $3.9$ nm CdSe core with a $2$-monolayer (ML) CdS shell QD contains a total of $1197$ Cd atoms, $483$ Se atoms, and $714$ S atoms. Phonon-mediated dephasing is treated phenomenologically by assigning dephasing rates $\gamma_S$ and $\gamma_B$ to excitonic state $|S\rangle$ and biexcitonic state $|B\rangle$, respectively. The dephasing rate is assumed to scale with both the reorganization energy and temperature.~\cite{lin2023theory,peng2026photoluminescence} Using the $X_2$ exciton as the reference state, the dephasing rate of excitonic state $|S\rangle$ is written as 
\begin{equation}
\gamma_S = \gamma_{X_2} \frac{\lambda_S}{\lambda_{X_2}}, 
\end{equation}
where $\lambda_S$ and $\lambda_{X_2}$ are the reorganization energies of state $|S\rangle$ and the reference state $|X_2\rangle$, respectively. The reorganization energy of excitonic state $S$ is defined as
\begin{equation}
\lambda_{S}=\frac{1}{2}\sum_{\alpha}\left(V_{SS}^{\alpha}/\omega_{\alpha}\right)^{2},
\end{equation}
where the exciton-phonon coupling strengths $V_{SS}^{\alpha}$ are obtained using semi-empirical pseudopotential methods,~\cite{jasrasaria2021interplay,jasrasaria2022simulations} and the vibrational frequencies $\omega_{\alpha}$ for mode $\alpha$ presented in the system are obtained by diagonalizing the dynamical matrix using the Stillinger-Weber potential.~\cite{jasrasaria2022simulations,zhou2013stillinger} Within a spatially uncorrelated biexciton model, the biexciton reorganization energy is approximated as the sum of the reorganization energies of its constituent single-exciton states. Under this framework, varying the system temperature allows the effective dephasing rate to span from several hundreds of ${\rm \mu eV}$ to several hundreds ${\rm meV}$.
The calculated reorganization energies of excitonic states with energies between approximately 1.9 and 2.2~eV are shown in Fig.~\ref{fig:QD_property}(a). High-lying excitonic states exhibit smaller reorganization energies compared with band-edge excitons. Fig.~\ref{fig:QD_property}(b) shows the phonon density of modes (PDOM):
\begin{equation}
\rho_{P}=\sum_{\alpha}\delta\left(\omega-\omega_{\alpha}\right),
\end{equation}
and the spectral density:
\begin{equation} J_S(\omega) = \frac{1}{2}\sum_{\alpha} \left(\frac{V_{SS}^{\alpha}}{\omega_\alpha}\right)^2 \delta(\omega-\omega_\alpha). 
\label{eq:spectral_density} 
\end{equation}
for the lowest bright excitonic state. The delta functions are broadened using a Gaussian function with a standard deviation of 0.3 meV.~\cite{ghosh2025atomistically}

\section{Bandwidth-Dependent Scaling in a Three-Level Model System}
\begin{figure}[t]
\includegraphics[width=17cm]{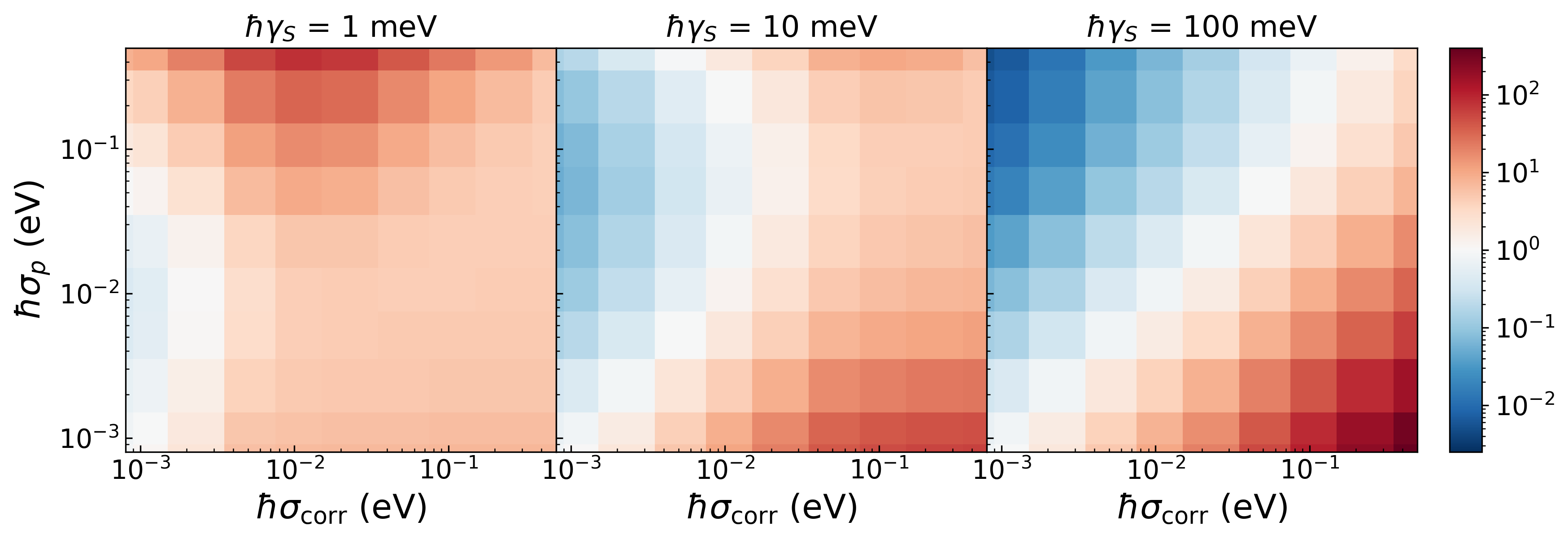}
\caption{Efficiency ratio $\eta$ in the three-level-system (3LS) model as a function
of pump bandwidth $\sigma_{p}$ and correlation bandwidth $\sigma_\mathrm{corr}$.
From left to right, the panels correspond to excitonic dephasing energies $\hbar\gamma_S=1$, $10$, and $100$~meV, respectively. The color scale indicates the value of $\eta$, defined as the quantum-to-classical enhancement of the biexciton-to-exciton population ratio. Values of $\eta>1$ indicate preferential enhancement of biexciton generation relative to single-exciton excitation under quantum light. \label{fig:3LS}}
\end{figure}

In Fig.~\ref{fig:3LS}, we plot the efficiency ratio $\eta$ as a function
of correlation bandwidth ($\sigma_\mathrm{corr}$) and pump bandwidth ($\sigma_{p}$),
corresponding to different excitonic dephasing energies $\hbar\gamma_{S}$.

\begin{figure}[t]
\includegraphics[width=17cm]{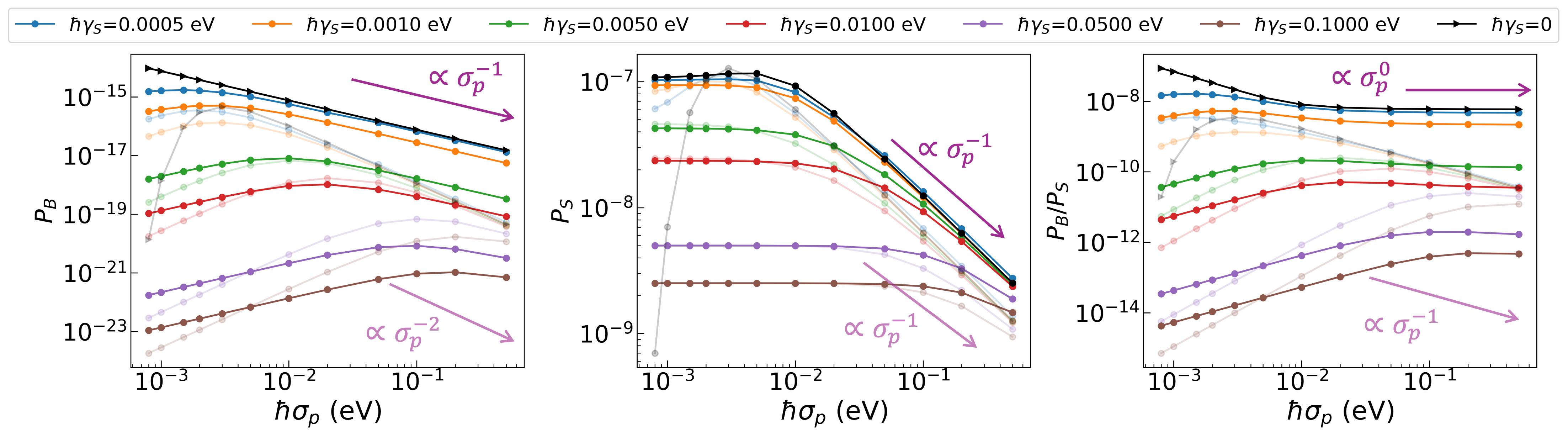}
\caption{Scaling of the biexciton population $P_B$ (left), exciton population $P_S$ (center), and biexciton-to-exciton population ratio $P_B/P_S$ (right) with the pump bandwidth $\hbar\sigma_p$ in the three-level-system model. The correlation bandwidth is fixed at $\hbar\sigma_{\mathrm{corr}}=6$~meV. Different colors correspond to the excitonic dephasing energies $\hbar\gamma_S$ indicated in the legend. Opaque curves show quantum-light results, while lighter curves of the same color show classical-light results. The black curves indicate the limiting behavior as $\gamma_S\rightarrow0$. The pink arrows and accompanying labels indicate the asymptotic power-law behavior in the large-bandwidth regime. Opaque arrows are for quantum light excitation results and transparent arrows are for classical light excitation results.\label{fig:scaling_p}}
\end{figure}

\begin{figure}
\includegraphics[width=17cm]{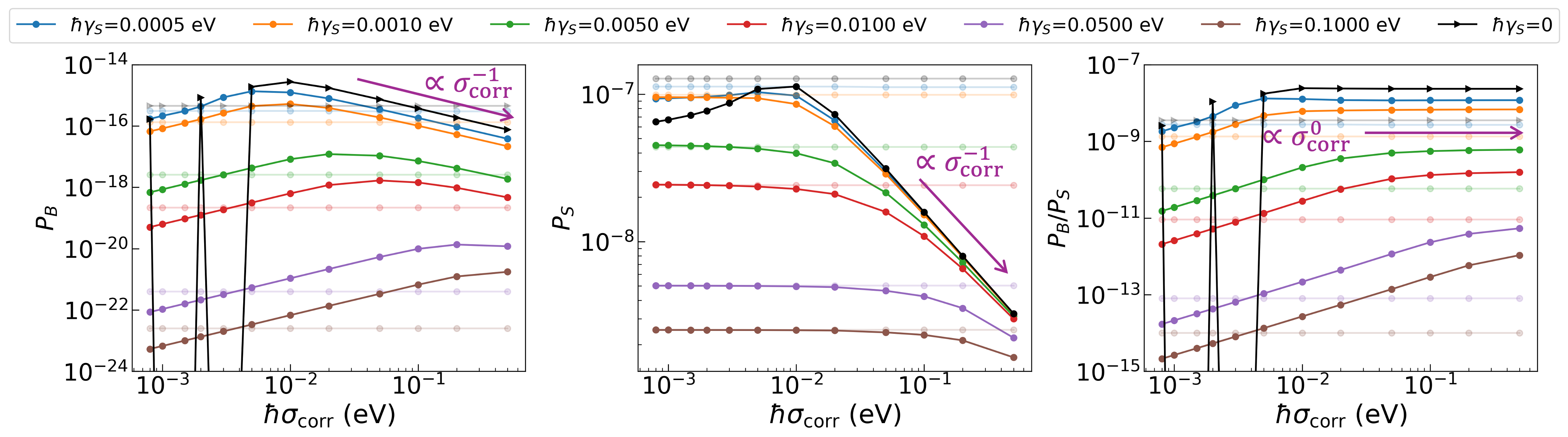}

\caption{Scaling of the biexciton population $P_B$ (left), exciton population $P_S$ (center), and biexciton-to-exciton population ratio $P_B/P_S$ (right) with the correlation bandwidth $\hbar\sigma_{\mathrm{corr}}$ in the three-level-system model. The pump bandwidth is fixed at $\hbar\sigma_p=3$~meV. Different colors correspond to the excitonic dephasing energies $\hbar\gamma_S$ indicated in the legend. Opaque curves show quantum-light results, while lighter curves of the same color show classical-light results. The black curves indicate the limiting behavior as $\gamma_S\rightarrow0$. The pink arrows and accompanying labels indicate the asymptotic power-law behavior in the large-bandwidth regime. Opaque arrows are for quantum light excitation results and transparent arrows are for classical light excitation results.
\label{fig:scaling_corr}}
\end{figure}

The advantage of the three-level-system (3LS) model is that it allows
the scaling behavior of excitation processes to be analyzed transparently.
In Fig.~\ref{fig:scaling_p} and Fig.~\ref{fig:scaling_corr}, we plot
the total biexciton population ($P_{B}$), total exciton population ($P_{S}$) and biexciton
population over exciton population ($P_{B}/P_{S}$) under both quantum light
(solid lines) and classical light (transparent lines) excitation. Different colors correspond to the excitonic dephasing energies $\hbar\gamma_S$ specified in the legends.
In our simulation, we define 
\begin{equation}
\frac{1}{\hbar\varepsilon_0 cA_0} = 10^{-11}\ \mathrm{D}^{-2},
\end{equation}
where $\varepsilon_0$ is the vacuum permittivity, $c$ is the speed of light, $A_0$ is the effective beam area, and $\mathrm{D}$ denotes Debye. This value corresponds to $A_0\approx4\times10^{-12}$~m$^2$.
In Fig.~\ref{fig:scaling_p}, correlation bandwidth is fixed at $6\text{ meV}$
for different dephasing rates $\hbar\gamma_{S}$ ranging from 0.0005 eV
to 0.1 eV. The limit $\gamma_{S}\rightarrow0$ is indicated by the black curves. The strong fluctuations originate from the oscillatory nature of the sinc function. In Fig. \ref{fig:scaling_corr}, pump bandwidth is
fixed at $3\text{ meV}$ for different dephasing rates $\gamma_{S}$.
The scaling relationship at large $\sigma_{p}$ and $\sigma_\mathrm{corr}$
is given in the plots denoted as pink arrow (opaque curves for quantum light
and transparent for classical light). When $\sigma_p$ or $\sigma_{\mathrm{corr}}$ becomes much larger than $\gamma_S$, power-law scaling emerges because the optical spectral amplitudes vary slowly over the material resonance. The scaling is then governed primarily by normalization of the spectral amplitudes. For classical light with a Gaussian
spectral envelope $\psi\left(\omega\right)$, $\left|\psi\left(\omega\right)\right|^{2}\propto\frac{1}{\sigma_{p}}$.
Consequently, 
\begin{equation}
\left[P_{B}\right]_{\text{Cl}}\propto\left|\psi\left(\omega\right)\psi\left(\omega\right)\right|^{2}\propto\frac{1}{\sigma_{p}^{2}},\text{ }\left[P_S\right]_{\text{Cl}}\propto\left|\psi\left(\omega\right)\right|^{2}\propto\frac{1}{\sigma_{p}},
\end{equation}
and they are independent of $\sigma_\mathrm{corr}$.
For quantum light described by a two-photon wavefunction $\Phi\left(\omega_{1},\omega_{2}\right)$, the
scaling depends on both pump and correlation bandwidths:
\begin{equation}
\left|\Phi\left(\omega_{1},\omega_{2}\right)\right|^{2}\propto\frac{1}{\sigma_{p}}\frac{1}{\sigma_\mathrm{corr}},
\end{equation}
where the additional scaling $1/\sigma_\mathrm{corr}$ arises from the normalization
of the sinc function. As a result, 
\begin{equation}
\left[P_{B}\right]_{\text{QL}}\propto\left|\Phi\left(\omega_{1},\omega_{2}\right)\right|^{2}\propto\frac{1}{\sigma_{p}}\frac{1}{\sigma_\mathrm{corr}},\text{ }\left[P_{S}\right]_{\text{QL}}\propto\left|\Phi\left(\omega_{1},\omega_{2}\right)\right|^{2}\propto\frac{1}{\sigma_{p}}\frac{1}{\sigma_\mathrm{corr}}.
\end{equation}
These asymptotic scaling relationships can be clearly observed in Fig. \ref{fig:scaling_p} and Fig.
\ref{fig:scaling_corr}.

\begin{figure}
\includegraphics[width=17cm]{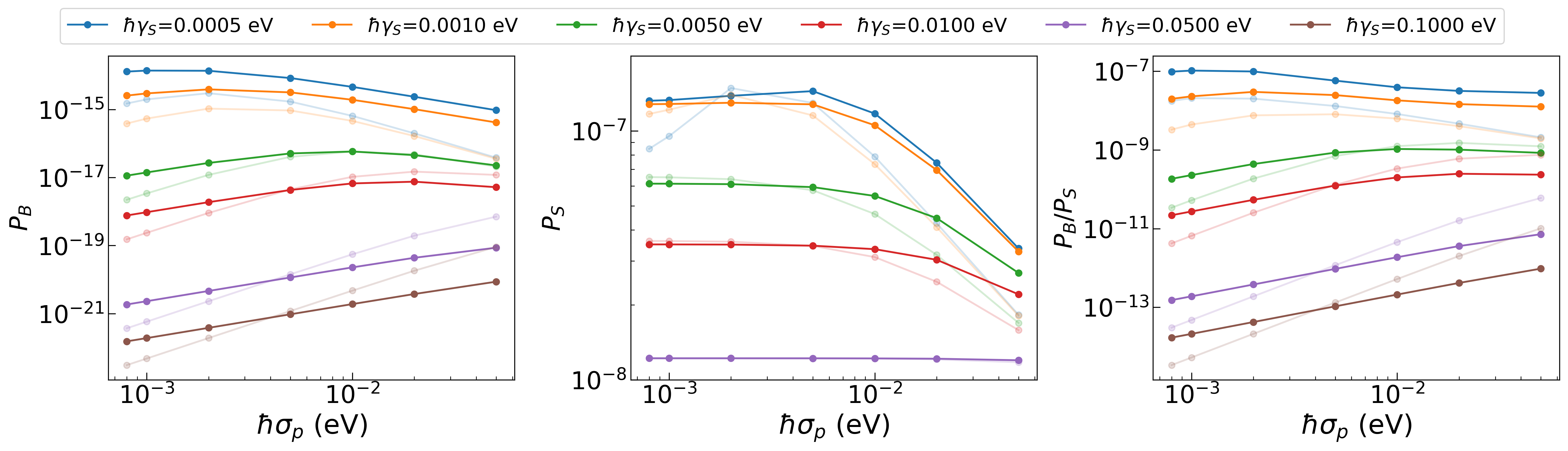}
\caption{Scaling of the biexciton population $P_B$ (left), exciton population $P_S$ (center), and biexciton-to-exciton population ratio $P_B/P_S$ (right) with the pump bandwidth $\hbar\sigma_p$ in the multi-level-system model. The correlation bandwidth is fixed at $\hbar\sigma_{\mathrm{corr}}=5$~meV. Different colors correspond to the excitonic dephasing energies $\hbar\gamma_S$ indicated in the legend. Opaque curves show quantum-light results, while lighter curves of the same color show classical-light results.
\label{fig:MLS_scaling_p}}
\end{figure}

\begin{figure}[t]
\includegraphics[width=17cm]{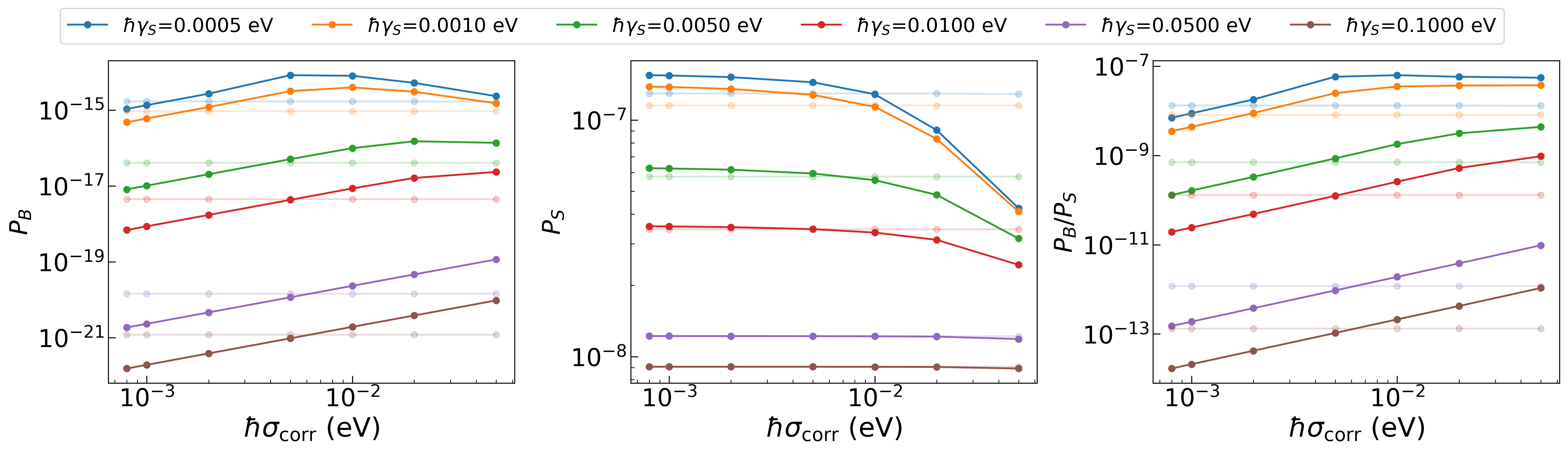}
\caption{Scaling of the biexciton population $P_B$ (left), exciton population $P_S$ (center), and biexciton-to-exciton population ratio $P_B/P_S$ (right) with the correlation bandwidth $\hbar\sigma_{\mathrm{corr}}$ in the multi-level-system model. The pump bandwidth is fixed at $\hbar\sigma_p=5$~meV. Different colors correspond to the excitonic dephasing energies $\hbar\gamma_S$ indicated in the legend. Opaque curves show quantum-light results, while lighter curves of the same color show classical-light results. 
\label{fig:MLS_scaling_corr}}
\end{figure}

\section*{Bandwidth Dependence in Realistic Multi-Level Quantum Dots}

To evaluate how the correlation-driven enhancement generalizes beyond the simplified three-level model, Figures~\ref{fig:MLS_scaling_p} and ~\ref{fig:MLS_scaling_corr} present the calculated biexciton population ($P_B$), single-exciton population ($P_S$), and the population ratio ($P_B/P_S$) for the full multi-level system (MLS) model representing a 3.9~nm CdSe core with a 2-ML CdS shell quantum dot.

In Figure~\ref{fig:MLS_scaling_p}, the correlation bandwidth is held constant at $\hbar\sigma_{\text{corr}} = 5$~meV while the pump bandwidth $\hbar\sigma_p$ is varied under different excitonic dephasing energies $\hbar\gamma_S$. Because expanding the bandwidth significantly would couple to higher-lying excitonic states beyond the finite manifold included in our model, both the correlation and pump bandwidths are constrained to $\le 50$~meV. Crucially, quantum light (opaque curves) consistently yields a higher biexciton-to-exciton population ratio $P_B/P_S$ than classical light (transparent curves) at optimal bandwidth across all dephasing regimes. This demonstrates that the quantum advantage in pathway selectivity remains robust within dense, realistic excitonic manifolds.

In Figure~\ref{fig:MLS_scaling_corr}, the pump bandwidth is fixed at $\hbar\sigma_p = 5$~meV while the correlation bandwidth $\hbar\sigma_{\text{corr}}$ is varied, directly controlling the biphoton entanglement time $\tau \sim 2\pi/\sigma_{\text{corr}}$. Unlike the single intermediate state in the 3LS model, the multi-level manifold contains multiple intermediate excitonic states $|S\rangle$ that participate in two-photon excitation. The coherent summation over these intermediate pathways gives rise to multi-pathway quantum interference. Despite this spectral complexity and phonon-induced dephasing, increasing $\sigma_{\text{corr}}$ drives $P_B/P_S$ toward a plateau, confirming that energy-time entanglement provides a durable mechanism for selective biexciton generation in colloidal nanocrystals.
\section{Derivation of Exciton and Biexciton Dynamics}
In the interaction picture, the time-evolution operator $U_I(t,t_0)$ satisfies the Schr\"odinger equation \begin{equation} \frac{\partial}{\partial t}U_I(t,t_0) = -\frac{i}{\hbar}V_I(t)U_I(t,t_0), \label{eq:interaction_evolution} \end{equation} where $t_0$ is the initial time, $V_I(t)$ is the interaction-picture light--matter coupling, and $\hbar$ is the reduced Planck constant.
The formal solution is given by the Dyson series:
\begin{equation}
U_{I}\left(t,t_{0}\right)=1+\left(-\frac{i}{\hbar}\right)\int_{t_{0}}^{t}dt_{1}V_{I}\left(t_{1}\right)+\left(-\frac{i}{\hbar}\right)^{2}\int_{t_{0}}^{t}dt_{2}\int_{t_{0}}^{t_{2}}dt_{1}V_{I}\left(t_{2}\right)V_{I}\left(t_{1}\right)+\cdots.
\end{equation}
The ordered time integrals enforce $t_0\leq t_1\leq t_2\leq t$, and the ellipsis denotes terms of third and higher order in $V_I$. Density matrix in interaction picture:
\begin{equation}
\rho_{I}\left(t\right)=U_{I}\left(t,t_{0}\right)\rho\left(t_{0}\right)U_{I}^{\dagger}\left(t,t_{0}\right).
\end{equation}
Here, $\rho(t_0)$ is the initial total density matrix of the quantum dot and photon field, and $\rho_I(t)$ is the corresponding interaction-picture density matrix.
Second order solution gives:

\begin{equation}
\begin{aligned}
& \rho_{I}^{(0)}(t)+\rho_{I}^{(1)}(t)+\rho_{I}^{(2)}(t) \\
&= U_{I}(t,t_{0})\rho(t_{0})U_{I}^{\dagger}(t,t_{0}) \\
&\approx\Bigg[1-\frac{i}{\hbar}\int_{t_{0}}^{t}dt_{1}\,V_{I}(t_{1})
+\left(-\frac{i}{\hbar}\right)^{2}\int_{t_{0}}^{t}dt_{2}\int_{t_{0}}^{t_{2}}dt_{1}
V_{I}(t_{2})V_{I}(t_{1})\Bigg]\quad  \\
&\times\rho(t_{0})\Bigg[1+\frac{i}{\hbar}\int_{t_{0}}^{t}dt_{1}\,V_{I}(t_{1})+\left(\frac{i}{\hbar}\right)^{2}\int_{t_{0}}^{t}dt_{2}\int_{t_{0}}^{t_{2}}dt_{1}
V_{I}(t_{1})V_{I}(t_{2})\Bigg] \\
&\approx\rho(t_{0})-\frac{i}{\hbar}\int_{t_{0}}^{t}dt_{1}
\Big[V_{I}(t_{1})\rho(t_{0})-\rho(t_{0})V_{I}(t_{1})\Big]\quad \\
&+\left(-\frac{i}{\hbar}\right)^{2}\int_{t_{0}}^{t}dt_{2}\int_{t_{0}}^{t_{2}}dt_{1}\Big[
V_{I}(t_{2})V_{I}(t_{1})\rho(t_{0})+\rho(t_{0})V_{I}(t_{1})V_{I}(t_{2}) \qquad\\
&-V_{I}(t_{1})\rho(t_{0})V_{I}(t_{2})- V_{I}(t_{2})\rho(t_{0})V_{I}(t_{1})\Big].
\end{aligned}
\end{equation}

To calculate the coherence between excitonic states, we retain the second-order contribution $\rho_I^{(2)}(t)$. The second-order term can be written as:
the formula for $\rho_{I}^{\left(2\right)}\left(t\right)$ as:
\begin{equation}
\rho_{I}^{(2)}\left(t\right) =\left(-\frac{i}{\hbar}\right)^{2}\int_{t_{0}}^{t}dt_{2}\int_{t_{0}}^{t_{2}}dt_{1}C^{(2)}\left(t_{1},t_{2}\right),
\end{equation}
\begin{equation}
C^{(2)}\left(t_{1},t_{2}\right)=V_{I}\left(t_{2}\right)V_{I}\left(t_{1}\right)\rho\left(t_{0}\right)+\rho\left(t_{0}\right)V_{I}\left(t_{1}\right)V_{I}\left(t_{2}\right)-V_{I}\left(t_{1}\right)\rho\left(t_{0}\right)V_{I}\left(t_{2}\right)-V_{I}\left(t_{2}\right)\rho\left(t_{0}\right)V_{I}\left(t_{1}\right).
\end{equation}
We take $t_0\rightarrow-\infty$ and assume the initially factorized state:
\begin{equation}
\rho(-\infty) = |g\rangle\langle g| \otimes |\Psi\rangle\langle\Psi|, 
\end{equation} where $|g\rangle$ is the quantum-dot ground state and $|\Psi\rangle$ is the incident biphoton state. For these different terms, only the last two terms contribute
to coherence and population at excitonic basis, so the reduced excitonic density-matrix element is obtained by projecting onto excitonic states $|S\rangle$ and $|S'\rangle$ and tracing over the photon degrees of freedom:
\begin{equation}
\begin{aligned}
\rho_{I,SS'} & =\text{Tr}_{f}\left[\left\langle S\right|\rho_{I}^{(2)}\left(t\right)\left|S'\right\rangle \right]\\
 & =\text{Tr}_{f}\left[\left\langle S\right|\left(-\frac{i}{\hbar}\right)^{2}\int_{-\infty}^{t}dt_{2}\int_{-\infty}^{t_{2}}dt_{1}\left\{ -V_{I}\left(t_{1}\right)\rho\left(t_{0}\right)V_{I}\left(t_{2}\right)-V_{I}\left(t_{2}\right)\rho\left(t_{0}\right)V_{I}\left(t_{1}\right)\right\} \left|S'\right\rangle \right]\\
 & =\frac{1}{\hbar^{2}}\text{Tr}_{f}\left[\int_{-\infty}^{t}dt_{2}\int_{-\infty}^{t_{2}}dt_{1}\left\langle S\right|V_{I}\left(t_{1}\right)\rho\left(t_{0}\right)V_{I}\left(t_{2}\right)\left|S'\right\rangle \right]_{ph}\\
 & +\frac{1}{\hbar^{2}}\text{Tr}_{f}\left[\int_{-\infty}^{t}dt_{2}\int_{-\infty}^{t_{2}}dt_{1}\left\langle S\right|V_{I}\left(t_{2}\right)\rho\left(t_{0}\right)V_{I}\left(t_{1}\right)\left|S'\right\rangle \right]_{Ph}\\
 & =\frac{1}{\hbar^{2}}\int_{-\infty}^{t}dt_{2}\int_{-\infty}^{t_{2}}dt_{1}\mu_{Sg}\mu_{S'g}e^{i\omega_{Sg}t_{1}}e^{-i\omega_{S'g}t_{2}}\left\langle \hat{E}^{(-)}\left(t_{2}\right)\hat{E}^{(+)}\left(t_{1}\right)\right\rangle _{f}\\
 & +\frac{1}{\hbar^{2}}\int_{-\infty}^{t}dt_{2}\int_{-\infty}^{t_{2}}dt_{1}\mu_{Sg}\mu_{S'g}e^{i\omega_{Sg}t_{2}}e^{-i\omega_{S'g}t_{1}}\left\langle \hat{E}^{(-)}\left(t_{1}\right)\hat{E}^{(+)}\left(t_{2}\right)\right\rangle _{f},
\end{aligned}
\end{equation}
where $\mathrm{Tr}_f$ denotes the trace over the photon field, $\mu_{Sg}$ and $\mu_{S'g}$ are the ground-to-exciton transition dipole moments, and $\omega_{Sg}$ and $\omega_{S'g}$ are the corresponding transition frequencies. The expectation value $\langle\cdots\rangle_f$ denotes an average over the initial biphoton state $|\Psi\rangle$.
\begin{equation}
\begin{aligned}
\left\langle \hat{E}^{(-)}\left(t_{1}\right)\hat{E}^{(+)}\left(t_{2}\right)\right\rangle _{f} & =\left\langle \Psi\right|\hat{E}^{(-)}\left(t_{1}\right)\hat{E}^{(+)}\left(t_{2}\right)\left|\Psi\right\rangle \\
 & =\int d\omega_{1}d\omega_{2}d\omega_{3}L\left(\omega_{3}\right)L\left(\omega_{1}\right)\left[\Phi^{*}\left(\omega_{3},\omega_{2}\right)\Phi\left(\omega_{1},\omega_{2}\right)+\Phi^{*}\left(\omega_{2},\omega_{3}\right)\Phi\left(\omega_{2},\omega_{1}\right)\right]\\
 &\times e^{i\omega_{3}t_{1}}e^{-i\omega_{1}t_{2}},
\end{aligned}
\end{equation}
\begin{equation}
\begin{aligned}
\left\langle \hat{E}^{(-)}\left(t_{2}\right)\hat{E}^{(+)}\left(t_{1}\right)\right\rangle _{f} & =\int d\omega_{1}d\omega_{2}d\omega_{3}L\left(\omega_{3}\right)L\left(\omega_{1}\right)\left[\Phi^{*}\left(\omega_{3},\omega_{2}\right)\Phi\left(\omega_{1},\omega_{2}\right)+\Phi^{*}\left(\omega_{2},\omega_{3}\right)\Phi\left(\omega_{2},\omega_{1}\right)\right]\\
&\times e^{i\omega_{3}t_{2}}e^{-i\omega_{1}t_{1}}.
\end{aligned}
\end{equation}
Here, \begin{equation} L(\omega) = \sqrt{\frac{\hbar\omega}{4\pi\varepsilon_0cA_0}} \end{equation} is the frequency-dependent field normalization, where $\varepsilon_0$, $c$, and $A_0$ are the vacuum permittivity, speed of light, and effective beam area, respectively.
In taking the steady-state limit, phenomenological dephasing rates $\gamma_S$ and $\gamma_{S'}$ are assigned to excitonic states $|S\rangle$ and $|S'\rangle$, respectively:

\begin{equation}
\begin{aligned}
\rho_{I,SS'} & =\lim_{t\rightarrow\infty}\text{Tr}_{f}\left[\left\langle S\right|\rho_{I}^{(2)}\left(t\right)\left|S'\right\rangle \right]\\
 & =\frac{\mu_{Sg}\mu_{S'g}}{2\hbar\varepsilon_{0}cA_{0}}\int d\omega_{1}\int d\omega_{2}\frac{-i}{\left(\omega_{Sg}-\omega_{1}-i\gamma_{S}\right)}\sqrt{\omega_{1}\left(\omega_{1}+\omega_{S'g}-\omega_{Sg}\right)}\\
 &\times\left[\Phi^{*}\left(\omega_{1}+\omega_{S'g}-\omega_{Sg},\omega_{2}\right)\Phi\left(\omega_{1},\omega_{2}\right)\right]\\
 & +\frac{\mu_{Sg}\mu_{S'g}}{2\hbar\varepsilon_{0}cA_{0}}\int d\omega_{1}\int d\omega_{2}\frac{-i}{\left(\omega_{Sg}-\omega_{2}-i\gamma_{S}\right)}\sqrt{\omega_{2}\left(\omega_{2}+\omega_{S'g}-\omega_{Sg}\right)}\\
 &\times\left[\Phi^{*}\left(\omega_{1},\omega_{2}+\omega_{S'g}-\omega_{Sg}\right)\Phi\left(\omega_{1},\omega_{2}\right)\right]\\
 & +\frac{\mu_{Sg}\mu_{S'g}}{2\hbar\varepsilon_{0}cA_{0}}\int d\omega_{1}\int d\omega_{2}\frac{i}{\left(\omega_{S'g}-\omega_{1}+i\gamma_{S'}\right)}\sqrt{\omega_{1}\left(\omega_{Sg}-\omega_{S'g}+\omega_{1}\right)}\\
&\times\left[\Phi^{*}\left(\omega_{1},\omega_{2}\right)\Phi\left(\omega_{Sg}-\omega_{S'g}+\omega_{1},\omega_{2}\right)\right]\\
 & +\frac{\mu_{Sg}\mu_{S'g}}{2\hbar\varepsilon_{0}cA_{0}}\int d\omega_{1}\int d\omega_{2}\frac{i}{\left(\omega_{S'g}-\omega_{2}+i\gamma_{S'}\right)}\sqrt{\omega_{2}\left(\omega_{Sg}-\omega_{S'g}+\omega_{2}\right)}\\
&\times\left[\Phi^{*}\left(\omega_{1},\omega_{2}\right)\Phi\left(\omega_{1},\omega_{Sg}-\omega_{S'g}+\omega_{2}\right)\right].
\end{aligned}
\end{equation}
For $S=S'$, the diagonal density-matrix element gives the exciton population:
\begin{equation}
\begin{aligned}
P_S & =\lim_{t\rightarrow\infty}\text{Tr}_{f}\left[\left\langle S\right|\rho_{I}^{(2)}\left(t\right)\left|S\right\rangle \right]\\
 & =\frac{\left|\mu_{Sg}\right|^{2}}{\hbar\varepsilon_{0}cA_{0}}\int d\omega_{1}d\omega_{2}\left[\frac{\gamma_{S}\omega_{1}}{\left(\omega_{Sg}-\omega_{1}\right)^{2}+\gamma_{S}^{2}}+\frac{\gamma_{S}\omega_{2}}{\left(\omega_{Sg}-\omega_{2}\right)^{2}+\gamma_{S}^{2}}\right]\left|\Phi\left(\omega_{1},\omega_{2}\right)\right|^{2}.
\end{aligned}
\end{equation}
Here, $P_S$ is the population of excitonic state $|S\rangle$, $\gamma_S$ is its dephasing rate, and $|\Phi(\omega_1,\omega_2)|^2$ is the biphoton joint spectral intensity.

Next, The biexciton population is evaluated to fourth order in the light--matter interaction. Equivalently, it may be written as the squared magnitude of the second-order transition amplitude:
\begin{equation}
\rho_{I,BB}\left(t\right)=\left|c_{B}\left(t\right)\right|^{2}=\left|-\frac{1}{\hbar^{2}}\int_{-\infty}^{t}dt_{2}\int_{-\infty}^{t_{2}}dt_{1}\left\langle 0,B\right|V_{I}\left(t_{2}\right)V_{I}\left(t_{1}\right)\left|\Psi,g\right\rangle \right|^{2}.
\end{equation}
Here, $|0,B\rangle\equiv|0\rangle_f\otimes|B\rangle$ denotes the photon vacuum and biexciton state, while $|\Psi,g\rangle\equiv|\Psi\rangle_f\otimes|g\rangle$ denotes the incident biphoton state and quantum-dot ground state.
The long expression here can be separated into two parts, molecular
part and photon part:

\begin{equation}
\begin{aligned}
c_{B}\left(t\right) & =-\frac{1}{\hbar^{2}}\int_{-\infty}^{t}dt_{2}\int_{-\infty}^{t_{2}}dt_{1}\left\langle 0,B\right|V_{I}\left(t_{2}\right)V_{I}\left(t_{1}\right)\left|\Psi,g\right\rangle \\
 & \approx-\frac{1}{\hbar^{2}}\int_{-\infty}^{t}dt_{2}\int_{-\infty}^{t_{2}}dt_{1}\left\langle B\right|\hat{d}^{(-)}\left(t_{2}\right)\hat{d}^{(-)}\left(t_{1}\right)\left|g\right\rangle \times\left\langle 0\right|\hat{E}^{(+)}\left(t_{2}\right)\hat{E}^{(+)}\left(t_{1}\right)\left|\Psi\right\rangle \\
 & =-\frac{1}{\hbar^{2}}\int_{0}^{t}dt_{2}\int_{0}^{t_{2}}dt_{1}\sum_{S}\mu_{BS}\mu_{Sg}e^{i\omega_{BS}t_{2}}e^{i\omega_{Sg}t_{1}}\\
 & \times\int d\omega'\int d\omega\int d\omega_{1}\int d\omega_{2}\Phi\left(\omega_{1},\omega_{2}\right)\frac{\hbar\sqrt{\omega\omega'}}{4\pi\varepsilon_{0}cA_{0}}e^{-i\omega t_{2}}e^{-i\omega't_{1}}\\
 & \times\left\langle 0\right|\left[\hat{a}_{1}\left(\omega'\right)+\hat{a}_{2}\left(\omega'\right)\right]\left[\hat{a}_{1}\left(\omega\right)+\hat{a}_{2}\left(\omega\right)\right]\hat{a}_{1}^{\dagger}\left(\omega_{1}\right)\hat{a}_{2}^{\dagger}\left(\omega_{2}\right)\left|0\right\rangle. 
\end{aligned}
\end{equation}
At steady state:
\begin{equation}
\begin{aligned}
P_{B} & =\lim_{t\rightarrow\infty}\left|c_{B}\left(t\right)\right|^{2}\\
 & =\left|\frac{\sum_{S}\mu_{BS}\mu_{Sg}}{2\hbar\varepsilon_{0}cA_{0}}\int_{0}^{\infty}d\omega_{2}\int_{0}^{\infty}d\omega_{1}\frac{\sqrt{\omega_{2}\omega_{1}}\Phi\left(\omega_{1},\omega_{2}\right)\gamma_{B}/\pi}{\left(\omega_{Bg}-\omega_{2}-\omega_{1}\right)^{2}+\gamma_{B}^{2}}\left[\frac{1}{\left(\omega_{1}-\omega_{Sg}+i\gamma_{S}\right)}+\frac{1}{\left(\omega_{2}-\omega_{Sg}+i\gamma_{S}\right)}\right]\right|^{2}.
\end{aligned}
\end{equation}
Here, $\omega_{Bg}=(E_B-E_g)/\hbar$ is the ground-to-biexciton transition frequency, $\gamma_B$ is the biexciton dephasing rate, and the sum over $S$ includes all intermediate excitonic states contributing to biexciton $|B\rangle$.

\section{Narrow Pump Bandwidth Limit}

We next evaluate the biexciton population in the narrow-pump-bandwidth limit, $\sigma_p\rightarrow0$, equivalently $T\rightarrow\infty$, where $T$ is the pump-pulse duration.
\begin{equation}
\begin{aligned}
c_{B}\left(t\right) & =-\frac{1}{\hbar^{2}}\int_{-\infty}^{t}dt_{2}\int_{-\infty}^{t_{2}}dt_{1}\left\langle 0,B\right|V_{I}\left(t_{2}\right)V_{I}\left(t_{1}\right)\left|\Psi,g\right\rangle \\
 & \approx-\frac{1}{\hbar^{2}}\frac{\hbar\sqrt{\omega_{2}^{0}\omega_{1}^{0}}}{4\pi\varepsilon_{0}cA_{0}}\sum_{S}\mu_{BS}\mu_{Sg}\int_{-\infty}^{\infty}dt_{2}\int_{0}^{+\infty}dt_{21}\int_{0}^{\infty}d\omega_{2}\int_{0}^{\infty}d\omega_{1}\times\\
 & \times\left[\Phi\left(\omega_{1},\omega_{2}\right)e^{i\omega_{Bg}t_{2}}e^{-i\omega_{2}t_{2}}e^{-i\omega_{1}\left(t_{2}-t_{21}\right)}e^{-i\omega_{Sg}t_{21}}+\Phi\left(\omega_{1},\omega_{2}\right)e^{i\omega_{Bg}t_{2}}e^{-i\omega_{1}t_{2}}e^{-i\omega_{2}\left(t_{2}-t_{21}\right)}e^{-i\omega_{Sg}t_{21}}\right].
\end{aligned}
\end{equation}
Recall the Fourier transformation of the joint spectral:
\begin{equation}
\Phi\left(\omega_{1},\omega_{2}\right)={\cal N}^{\text{QM}}e^{-\left(\omega_{1}+\omega_{2}-\omega_{1}^{0}-\omega_{2}^{0}\right)^{2}T^{2}}\text{Sinc}\left(\left[\omega_{1}-\omega_{2}-\omega_{1}^{0}+\omega_{2}^{0}\right]\tau\right),
\end{equation}
\begin{equation}
\begin{aligned}
\Phi\left(t_{1},t_{2}\right) & =\mathcal{F}\left[\Phi\left(\omega_{1},\omega_{2}\right)\right]=\int_{-\infty}^{\infty}d\omega_{1}d\omega_{2}e^{-i\omega_{1}t_{1}}e^{-i\omega_{2}t_{2}}\Phi\left(\omega_{1},\omega_{2}\right)\\
 & ={\cal N}_{t}^{{\rm QM}}e^{-\left(t_{1}-T_{d}+t_{2}\right)^{2}/16T^{2}}\Pi_{\tau}\left(t_{1}-T_{d}-t_{2}\right)e^{-i\omega_{1}^{0}t_{1}}e^{-i\omega_{2}^{0}t_{2}},
\end{aligned}
\end{equation}
where, $\mathcal N^{\mathrm{QM}}$ is the frequency-domain normalization constant, $\omega_1^0$ and $\omega_2^0$ are the central photon frequencies, $T$ is the pump-pulse duration, $\tau$ is the biphoton entanglement time, $\mathcal F$ denotes the two-dimensional Fourier transform, $\mathcal N_t^{\mathrm{QM}}$ is the time-domain normalization constant satisfying $\int dt_1 dt_2 |\Phi(t_1,t_2)|^2=1$, $T_d$ is the relative delay between the two photon modes (we set $T_d=0$ in the calculations presented in this work), and $\Pi_\tau(t)$ is a rectangular temporal function whose width is determined by the entanglement time $\tau$:
\begin{equation}
\Pi_\tau(t)= \begin{cases} 1, & |t|\leq\tau/2,\\ 0, & |t|>\tau/2. 
\end{cases} 
\end{equation}
So that at steady state, biexciton population $P_B$ under narrow band excitation can be written as:
\begin{equation}
\begin{aligned}
P_B
= \Bigg|
{\cal N}_{t}^{\mathrm{QM}}
\frac{\sqrt{\omega_{2}^{0}\omega_{1}^{0}}}{2\hbar\varepsilon_{0}cA_{0}}
\sum_{S}\mu_{BS}\mu_{Sg}
\frac{\gamma_{B}/\pi}
{\left(\omega_{Bg}-\omega_{2}^{0}-\omega_{1}^{0}\right)^{2}+\gamma_{B}^{2}}
\\
\times
\left[
\frac{e^{-i(\omega_{Sg}-\omega_{1}^{0}-i\gamma_{S})\tau}-1}
{\omega_{Sg}-\omega_{1}^{0}-i\gamma_{S}}
+
\frac{e^{-i(\omega_{Sg}-\omega_{2}^{0}-i\gamma_{S})\tau}-1}
{\omega_{Sg}-\omega_{2}^{0}-i\gamma_{S}}
\right]
\Bigg|^{2}.
\end{aligned}
\end{equation}
The two terms in brackets correspond to the two possible photon-absorption orderings. Their coherent sum retains the relative phase accumulated through each intermediate excitonic state during the entanglement time $\tau$.
\bibliography{reference}